\documentstyle[12pt,epsfig]{article}
\newcommand{\be}{\begin{equation}}
\newcommand{\ee}{\end{equation}}
\newcommand{\beq}{\begin{equation}}
\newcommand{\eeq}{\end{equation}}
\newcommand{\ba}{\begin{eqnarray}}
\newcommand{\ea}{\end{eqnarray}}
\newcommand{\bea}{\begin{eqnarray}}
\newcommand{\eea}{\end{eqnarray}}
\newcommand{\nn}{\nonumber}
\begin{document}
\baselineskip=15.5pt \pagestyle{plain} \setcounter{page}{1}
%
%--------+---------+---------+---------+---------+---------+---------+
%--------+---------+---------+---------+---------+---------+---------+
\begin{titlepage}

\vskip 0.8cm

\begin{center}

{\Large \bf Deep inelastic scattering structure functions of \\
\vskip 0.2cm holographic spin-1 hadrons with $N_f \geq 1$}

\vskip 1.cm

{\large {{\bf Ezequiel Koile}$^{a,}${\footnote{\tt
koile@fisica.unlp.edu.ar}}, {\bf Sebastian
Macaluso}$^{b,}${\footnote{\tt macaluso@physics.rutgers.edu }} {\bf
and Martin Schvellinger}$^{a,}${\footnote{\tt
martin@fisica.unlp.edu.ar}}}}

\vskip 1.cm

$^{a}${\it IFLP-CCT-La Plata, CONICET and Departamento  de
F\'{\i}sica,  Universidad Nacional de La Plata.  Calle 49 y 115,
C.C. 67, (1900) La Plata,  Buenos Aires,
Argentina.} \\

\vskip 0.5cm

$^{b}${\it Department of Physics and Astronomy, Rutgers University,
Piscataway, NJ 08854-8019, USA.} \\

\vspace{1.cm}

{\bf Abstract}

\vspace{1.cm}

\end{center}

Two-point current correlation functions of the large $N$ limit of
supersymmetric and non-supersymmetric Yang-Mills theories at strong
coupling are investigated in terms of their string theory dual
models with quenched flavors. We consider non-Abelian global
symmetry currents, which allow one to investigate vector mesons with
$N_f > 1$. From the correlation functions we construct the deep
inelastic scattering hadronic tensor of spin-one mesons, obtaining
the corresponding eight structure functions for polarized vector
mesons. We obtain several relations among the structure functions.
Relations among some of their moments are also derived. Aspects of
the sub-leading contributions in the $1/N$ and $N_f/N$ expansions
are discussed. At leading order we find a universal behavior of the
hadronic structure functions.

\noindent

\end{titlepage}

\newpage

{\small \tableofcontents}

\newpage

%--------------------------------------------------------------------
\section{Introduction}\label{Introduction}
%--------------------------------------------------------------------

Two-point current correlation functions are relevant for the
calculation of important observables of quantum field theories. In
particular, for confining gauge theories they allow one to construct
the so-called hadronic tensor of deep inelastic scattering (DIS)
processes, which is an invaluable tool to extract fundamental
information about the structure of hadrons. When considering a DIS
process the idea is that a lepton is scattered from a hadron, being
the interaction mediated by a virtual photon exchanged from the
lepton to the hadron. The process is called {\it inclusive} since
only the scattered lepton is measured, while the hadronic final
state is not. The differential cross section of DIS is given by the
contraction of a leptonic tensor, which is obtained from Quantum
Electrodynamics, and a hadronic tensor, $W_{\mu\nu}$, which carries
the information about the strong interaction. By using the optical
theorem in quantum field theory the hadronic tensor can be written
in terms of the vacuum expectation value of the product of two
currents. $W_{\mu\nu}$ has a Lorentz tensor decomposition which
depends on the spin of the hadron, and it can be expressed as a sum
of several terms. In addition, there are functions multiplying each
of these terms. These structure functions, like the tensor
$W_{\mu\nu}$, should in principle be derived from QCD. However, the
non-perturbative character of QCD makes it extremely difficult to
obtain such functions in that way. The structure of the hadronic
tensor lies on the two-point current correlation functions, which
are affected by the non-perturbative nature of soft processes of
QCD.

On the other hand, the gauge/string duality provides holographic
dual models which can actually be used to calculate the structure
functions of hadrons derived from such models, in terms of two-point
current correlation functions. This is so because within this
duality the non-perturbative regime of the quantum field theory
corresponds to the perturbative regime of the holographic string
theory dual model. In this paper we investigate properties of
two-point current correlation functions, and therefore the DIS
hadronic tensor, using different holographic string theory dual
models with flavors in the fundamental representation of the gauge
group, and within the quenched approximation. The hadrons we
consider are mesons. Notice that these mesons are not exactly those
of QCD because at present there is not any holographic dual model
which accounts for all the properties of QCD, even in the large $N$
limit. However, it is very interesting to be able to explore their
internal structure, since it could manifest a universal character,
which obviously is inherent to the two-point current correlation
functions. In fact we find such a universal behavior. Particularly,
we are interested in the study of two-point correlation functions of
non-Abelian symmetry currents, which allows one to describe hadrons
with different flavor content.

Polchinski and Strassler proposed a model for the holographic dual
description of DIS of confining gauge theories
\cite{Polchinski:2002jw} that we briefly describe below. They
calculated hadronic structure functions for the Bjorken parameter
$x$ of order one within the supergravity approximation. They also
considered a small-$x$ calculation by using a dual string theory
analysis. Their approach is in the large $N$ limit of confining
supersymmetric Yang-Mills theories in four dimensions, such as
certain deformations of ${\cal{N}}=4$ SYM, from which they study DIS
from glueballs and spin-$\frac{1}{2}$ hadrons. The gauge theories
studied in \cite{Polchinski:2002jw} are UV conformal or nearly
conformal, which makes the dual string theory defined on a
background of the type AdS$_5 \times {\cal {M}}_5$, being ${\cal
{M}}_5$ a compact five-dimensional Einstein manifold. Thus, this is
a solution of type IIB supergravity whose metric can be written as
\begin{equation}\label{pol1}
ds^{2}=\frac{\rho^{2}}{R^{2}} \, \eta_{\mu\nu} \, dy^{\mu} \,
dy^{\nu} + \frac{R^{2}}{\rho^{2}} \, d\rho^{2} + R^{2} \,
\widehat{ds}^2_{{\cal {M}}_5}\, ,
\end{equation}
where the AdS$_5$ radius is $R=(4\pi g_s N)^{1/4}\alpha'^{1/2}$ when
${\cal {M}}_5$ is $S^5$. The four-dimensional gauge field theory
coordinates are identified with $y^{\mu}$, while $\rho$ is the
holographic radial coordinate related to the dual quantum field
theory energy scale. Up to powers of the 't Hooft coupling $\lambda
= g_{YM}^2 N \equiv 4 \pi g_s N$, the ten-dimensional energy scale
is given by $R^{-1}$, where the string coupling is denoted by $g_s$.
Thus, the four-dimensional energy is given by
\begin{equation}\label{pol2}
E^{(4)}\sim\frac{\rho}{R^{2}}\, .
\end{equation}
In the large $N$ limit of confining gauge theories, the geometry of
the holographic dual model whose metric is given by Eq.(\ref{pol1})
must be modified at a radius corresponding to $\rho \sim
\rho_{0}=\Lambda R^{2}$. Notice the presence of a confinement scale
$\Lambda$. It is worth mentioning that the dynamics of interest for
$q \gg \Lambda$ lies on the region where $\rho_{int} \sim q R^{2}
\gg \rho_{0}$, where $\rho_{int}$ denotes the bulk region where the
relevant interaction occurs. Within this region the conformal metric
(\ref{pol1}) can be used. Thus, it is possible to calculate the dual
of the matrix element of the $T^{\mu\nu}$ tensor, which as we shall
explain in section 2, is related to the hadronic tensor. As
commented before, by using the optical theorem we can write its
imaginary part as
\begin{eqnarray}\label{pol3}
\textmd{Im} \:T^{\mu\nu} &=& \pi\sum_{P_{X},X}\langle
P,{\cal{Q}}|J^{\nu}(0)|P_{X},X\rangle\langle
P_{X},X|\widetilde{J}^{\mu}(q)|P,{\cal{Q}}\rangle \nonumber\\
&=& 2\pi^{2}\sum_{X}\delta(M_{X}^{2}+[P+q]^{2})\langle P,
{\cal{Q}}|J^{\nu}(0)|P+q, X\rangle \langle
P+q,X|J^{\mu}(0)|P,{\cal{Q}}\rangle\, ,
\end{eqnarray}
which has been written in terms of the hadron ($P_{\mu}$) and
virtual photon ($q_{\mu}$) momenta, and the currents $J_{\mu}$.
There is a sum over intermediate states $X$ with mass $M_X$. Notice
that $\eta^{\mu\nu}$ raises their Lorentz indices which are
four-dimensional ones.

In the large $N$ limit of the gauge theory only single hadron states
will contribute. If $-P^{2} \ll q^{2}$, \emph{i.e.} $|t|\ll 1$, then
in the $s$-channel we can approximate
\begin{eqnarray}\label{pol20}
s=-(P+q)^{2}\simeq q^{2}\bigg(\frac{1}{x}-1 \bigg)\, ,
\end{eqnarray}
where we have used the Bjorken variable
\begin{equation}
x\equiv -\frac{q^2}{2 P \cdot q} \, ,
\end{equation}
and also
\begin{equation}
t \equiv \frac{P^2}{q^2} \,.
\end{equation}
The condition $-P^{2} \ll q^{2}$ is equivalent to $|t|\ll 1$.
On the other hand, in ten dimensions the scale $\widetilde{s}$ is
set by the relation
\begin{eqnarray}\label{pol21}
\widetilde{s} = -g^{MN}P_{X,M}P_{X,N}\leq
-g^{\mu\nu}(P+q)_{\mu}(P+q)_{\nu}
\sim\frac{R^{2}}{\rho^{2}_{int}}q^{2}\bigg(\frac{1}{x}-1
\bigg)=\frac{\bigg(x^{-1}-1 \bigg)}{\alpha'(4\pi g_s N)^{1/2}} \, .
\end{eqnarray}
The 't Hooft parameter appears in the denominator, so if $(g_s
N)^{-1/2} \ll x<1$ we have $\alpha' \widetilde{s} \ll 1$. Therefore,
in this limit only massless string states are produced, and we are
dealing with a purely supergravity process \cite{Polchinski:2002jw}.
Through this work we assume the Bjorken variable to be within the
kinematical regime where the supergravity approximation is reliable.

One can describe the DIS process from the bulk theory perspective.
The idea is that within the four-dimensional boundary theory we
consider the two-point function of two global symmetry currents
inside the hadron. So, let us consider the effect of the insertion
of a current operator at the boundary of the AdS$_5$ space-time.
This leads to a perturbation on the boundary condition of a bulk
gauge field. This perturbation produces a non-normalizable mode
propagating in the bulk \cite{Gubser:1998bc,Witten:1998qj}. In order
to find this mode we should look at the isometry group of the
manifold ${\cal {M}}_5$, which corresponds to an $R$-symmetry group
on the boundary field theory. If one takes a $U(1)_R$ subgroup, the
associated $R$-symmetry current can be identified with the
electromagnetic current inside the hadron. Notice that for the
global symmetry group, which corresponds to the isometry of ${\cal
{M}}_5$, there is a Killing vector $\upsilon_{j}$ which produces the
non-normalizable mode of a Kaluza-Klein gauge field $A_{m}(y,r)$.
Therefore, the metric perturbation induced by the $R$-symmetry
current operator is
\begin{eqnarray}\label{pol5}
\delta g_{mj}=A_{m}(y,r) \, \upsilon_{j}(\Omega)\, .
\end{eqnarray}
This mode $A_{m}(y,r)$ propagates in the bulk and couples to a bulk
field which is dual to a certain quantum field theory state. For
instance when considering glueballs, the holographic dual field in
\cite{Polchinski:2002jw} corresponds to the dilaton. Thus, the
incoming bulk dilaton field $\Phi_i$ couples to the bulk
$U(1)$-gauge field $A_\mu$ (induced by a current operator inserted
at the boundary) and to another dilaton $\Phi_X$, which represents
an intermediate hadronic state\footnote{The corresponding tree level
supergravity Witten's diagram can be seen in figure 1, which
describes a forward Compton scattering. Notice that in that figure
solid lines indicate mesons.}. The intermediate state propagates in
the bulk and couples to an outgoing dilaton $\Phi_f$ (corresponding
to the final hadronic state) and a gauge field $A_\nu$ in the bulk
which comes from the insertion of a second boundary theory current
operator. This is nothing but a holographic dual version of the
quantum field theory optical theorem. This can be generalized to
other situations, namely mesons including flavors in the fundamental
representation of the gauge group. In this case $A_{m}(y,r)$ in the
bulk couples to either scalar or vector fluctuations of flavor probe
branes, and the two-point functions which lead to the hadronic
tensor correspond to non-Abelian global symmetry currents.

Another interesting issue is related to the role of the sub-leading
corrections to the Operator Product Expansion (OPE) of two symmetry
currents. From this, the moments of the structure functions can be
obtained. These moments have different kind of contributions to the
$1/N$ expansion, {\it i.e.} while at weak coupling single-trace
twist-two operators dominate the expansion, at strong coupling
double-trace operators become relevant \cite{Polchinski:2002jw}.

We can summarize our main results as follows. We have performed a
detailed analysis of the structure of the two-point correlation
functions of generic symmetry currents at strong coupling,
associated with flavors in the fundamental representation of the
gauge group, in the quenched approximation, in terms of the
corresponding holographic string theory dual description. This
includes the large $N$ limit of supersymmetric and
non-supersymmetric Yang-Mills theories in four dimensions. In
particular, we have explicitly investigated the cases of the
D3D7-brane, the D4D8$\mathrm{\overline{D8}}$-brane, and the
D4D6$\mathrm{\overline{D6}}$-brane systems. We would like to
emphasize that we have found a {\it universal} structure of the
two-point correlation functions of generic global symmetry currents
at strong coupling. For each holographic dual model we have found
that the two-point correlation functions of non-Abelian ($N_f > 1$)
global symmetry currents can generically be written as the product
of a constant, which depends on the particular Dp-brane model, times
flavor preserving Kronecker deltas multiplying the corresponding
Abelian ($N_f=1$) result for the same Dp-brane model. We have
obtained a universal factorization of the two-point correlation
functions for non-Abelian symmetry currents in a model-dependent
factor times a model-independent one. More precisely, we should
stress that these results strictly hold in the large $N$ limit, {\it
i.e.} to leading order in the $1/N$ expansion. Sub-leading
corrections in this expansion would likely induce some
modifications, obviously negligible in the large $N$ limit. The
model-dependent and model-independent factorization has already been
seen for the two-point functions of Abelian symmetry currents in our
previous paper \cite{Koile:2011aa}. This factorization comes from
the structure of the flavored holographic dual model in the probe
approximation, where the probe Dp-brane action is taken to be the
non-Abelian version of the Dirac-Born-Infeld action
\cite{Tseytlin:1997csa}. Thus, in general we can write the
$W^{\mu\nu}_{(a)}$ tensor for a holographic dual model corresponding
to a certain gauge field theory in the large-$N$ limit as
\begin{equation}
W^{\mu\nu}_{(a)}=A_{(a, b)}(x) \, W^{\mu\nu}_{(b)} \, ,
\end{equation}
for models $(a)$ and $(b)$, where $A_{(a, b)}$ is a conversion
factor which depends on the pair of Dp-brane models considered. This
allows one to write the corresponding structure functions
$F_i^{(a)}(x, t)$, where subindex $i$ indicates the $i$-th structure
function for every meson in each particular model, as
\begin{equation}
F^{(a)}_i(x, t) = A_{(a, b)}(x) \, F^{(b)}_i(x, t) \, .
\end{equation}
Besides, we have found that a modified version of the Callan-Gross
relation is satisfied by the class of flavored holographic dual
models we have investigated, when the parameter $t \rightarrow 0$.
We have obtained new relations between structure functions for the
$N_f>1$ case within each particular model. This confirms our results
for $N_f=1$ given in \cite{Koile:2011aa}. This suggests that these
relations among structure functions are generic and, therefore it
may indicate that they hold for any confining gauge theory in the
appropriate kinematical regime. In addition, we have shown that all
the moments of certain structure functions satisfy the corresponding
inequalities derived from unitarity, as expected
\cite{Manohar:1992tz}.

A very interesting aspect of the present work is that we have
investigated the $1/N$ and $N_f/N$ contributions to the leading
order calculations of the hadronic tensor, from the supergravity
dual model point of view. Particularly, we have focused on the
structure of the relevant Lagrangians and Witten's diagrams. Indeed,
we have derived all relevant Lagrangians. On the other hand,
although we have not calculated these Witten's diagrams explicitly,
we have discussed how they arise from supergravity. We have pointed
out that the $1/N$ and $N_f/N$ expansions of the Witten's diagrams
correspond to analogous expansions in the dual quantum field theory.
We also have shown how these Witten's diagrams are suppressed by
$1/N^2$ and $N_f/N$ powers, respectively, in the supergravity dual
models.

This paper essentially contains two parts. The first one, which
includes sections 2, 3 and 4, develops a non-trivial generalization
of our results of reference \cite{Koile:2011aa} when the number of
flavors is larger than one, but still within the quenched
approximation. In section 3 we begin with a general background
metric, which includes the two cases studied in \cite{Koile:2011aa},
as well as the D4D6$\mathrm{\overline{D6}}$-brane system
\cite{Kruczenski:2003uq}. We calculate the structure functions for
scalar and vector mesons. In section 4, we extend this approach to
study flavored vector mesons, which is done in the gravity dual
theory by adding $N_f$ flavor probe Dp-branes, with $1 < N_f \ll N$.
The second part is introduced in section 5 and it contains very
interesting new results about the $1/N$ expansion. We have discussed
results corresponding to a DIS process where the lepton is scattered
from an entire hadron, which becomes excited but is not fragmented.
Beyond it, in section 5 we have considered the $1/N$ and $N_f/N$
expansions. It would be very interesting to investigate the effects
of the back-reaction of the probe Dp-branes on the background beyond
the probe approximation. Another aspect we have not considered
concerns the kinematic regime where the Bjorken parameter is very
small, whose holographic dual description goes beyond pure
supergravity. In section 6 we carry out a discussion of our results.
Two appendices are included to account for details of expressions
commented in the main text, and in order to include explicit results
of the two-point current correlations functions for the
D4D6$\mathrm{\overline{D6}}$-brane model.

\newpage

%--------------------------------------------------------------------------
\section{Two-point current correlation functions and DIS}\label{Kinematics}
%--------------------------------------------------------------------------

In what follows we adopt the conventions of Manohar
\cite{Manohar:1992tz}, except for the Minkowski metric, which we
define as being mostly plus. A brief review of the relevant ideas
and definitions for the present work can be found in our previous
paper \cite{Koile:2011aa}. A more detailed derivation of DIS
structure functions is available in references \cite{Manohar:1992tz}
and \cite{Hoodbhoy:1988am}.

We consider an incoming lepton beam with four-momentum $k^\mu$ (with
$k^0\equiv E$) which will be scattered from a fixed hadronic target.
The four-momentum of the scattered lepton $k'^\mu$ (with $k'^0\equiv
E'$) is measured, but the final hadronic state called $X$ is not.
The lepton and the initial hadronic state exchange a virtual photon
with four-momentum $q^{\mu}$. Thus, this virtual photon is able to
probe the hadron structure at distances as small as
$1/\sqrt{q^{2}}$.

The DIS differential cross section can be written as
\begin{equation}\label{DIS7}
\frac{d^{2}\sigma}{dE'd\Omega}=\frac{e^{4}}{16\pi^{2}q^{4}}
\frac{E'}{ME}l^{\mu\nu}W_{\mu\nu}(P,q)_{h'\,h}\, ,
\end{equation}
where we have defined the leptonic tensor as follows
\begin{equation}\label{DIS3.1}
l^{\mu\nu}=\sum_{final \ spin}\langle k'|J_l^{\nu}(0)|k,s_l\rangle
\langle k,s_l|J_l^{\mu}(0)|k'\rangle\, ,
\end{equation}
which for a spin-$\frac{1}{2}$ lepton becomes
\begin{equation}\label{DIS3.2}
l^{\mu\nu}=2 \, [k^\mu k'^\nu + k^\nu k'^\mu -\eta^{\mu\nu} (k\cdot
k'-m_l^2) - i \, \epsilon^{\mu\nu\alpha\beta} \, q_\alpha \,
s_{l\beta}] \, ,
\end{equation}
being $m_{l}$ the lepton mass. In addition, the hadronic tensor is
\begin{equation}\label{DIS4}
W_{\mu\nu}(P,q)_{h'\,h} = \frac{1}{4\pi} \int d^{4}x \, e^{iq.x} \,
\langle P,h'|[J_{\mu}(x),J_{\nu}(0)]|P,h\rangle\,,
\end{equation}
where $P^\mu$ and $P_X^\mu$ denote the hadronic initial and final
momenta, $h$ and $h'$ are the polarizations of the initial and final
hadronic states, and $M^2=-P^2$ and $M_X^2=-P_X^2$ are the initial
and final hadronic squared masses, respectively. The hadronic tensor
can be recast in terms of its structure functions. In fact, the
so-called partonic distribution functions, which can be calculated
from the structure functions, give the probability that a hadron
contains a given constituent with a given fraction $x$ of its total
momentum. Due to the non-perturbative character of QCD, since the
partonic distribution functions depend on soft QCD dynamics, they
cannot be extracted perturbatively.

In the case of hadrons composed by massless partons, the probability
of finding a parton with a momentum $x P^\mu$ is given by the
distribution function $f(x, q^{2})$. In the case of free partons
this function leads to the Bjorken scaling, which is not actually
true for QCD since it is not a free field theory. Notice that the
hadronic structure functions are dimensionless functions of $P^{2}$,
$P \cdot q$ and $q^{2}$. It is usual to write their functional
dependence in terms of $t$ and $x$ variables described in the
introduction, with $0 < x \le 1$ and $t \le 0$. The structure
functions are obtained from the most general Lorentz decomposition
of the hadronic tensor $W_{\mu\nu}$, satisfying parity invariance,
time reversal symmetry, and invariance under translations.

The most general form for spin-zero targets is
\cite{Polchinski:2002jw}
\begin{equation}\label{DIS16sca}
W_{\mu\nu}^{scalar} = F_{1}\bigg(\eta_{\mu\nu}-\frac{q_\mu
q_\nu}{q^2}\bigg)-\frac{F_{2}}{P \cdot
q}\bigg(P_{\mu}+\frac{q_\nu}{2x}\bigg)\bigg(P_{\nu}+\frac{q_\nu}{2x}\bigg)
.
\end{equation}
After contracting with $l^{\mu\nu}$, terms containing $q_\mu$ and
$q_\nu$ vanish. Therefore, we can just neglect these terms from the
beginning obtaining a simpler expression
\begin{equation}\label{DIS16scasimple}
W_{\mu\nu}^{scalar}= F_{1} \, \eta_{\mu\nu} - \frac{F_{2}}{P \cdot
q}P_{\mu}P_{\nu}.
\end{equation}
For spin-one targets, on the other hand, the full general form of
the hadronic tensor is \cite{Hoodbhoy:1988am}
\begin{eqnarray}\label{DIS16}
W_{\mu\nu}^{vector} &=& F_{1} \, \eta_{\mu\nu}-\frac{F_{2}}{P \cdot
q}P_{\mu}P_{\nu}+b_{1}r_{\mu\nu}-\frac{b_{2}}{6}(s_{\mu\nu}+t_{\mu\nu}+
u_{\mu\nu}) -\frac{b_{3}}{2}(s_{\mu\nu}-u_{\mu\nu})\nonumber\\ & &
-\frac{b_{4}}{2}(s_{\mu\nu}-t_{\mu\nu}) -\frac{i \, g_{1}}{P \cdot
q} \, \epsilon_{\mu\nu\lambda\sigma} \, q^{\lambda} \, s^{\sigma}
-\frac{i \, g_{2}}{(P \cdot q)^{2}} \,
\epsilon_{\mu\nu\lambda\sigma} \, q^{\lambda}
\, (P \cdot q \:s^{\sigma}-s \cdot q \:P^{\sigma})\, , \nonumber\\
\end{eqnarray}
where we have already omitted terms proportional to $q_\mu$ and
$q_\nu$, as explained before. Functions $r_{\mu\nu}$, $s_{\mu\nu}$,
$t_{\mu\nu}$, $u_{\mu\nu}$ and $s^\sigma$, which depend on the
hadron polarization, on the hadron and virtual photon momenta, and
on the $t$ and $x$ variables, are defined in appendix A.

DIS amplitudes can be obtained from the imaginary part of the
forward Compton scattering amplitudes. Thus, it is possible to
define the tensor
\begin{equation}\label{DIS50}
T_{\mu\nu}=i \, \langle P, {\cal {Q}}|{\widehat{T}}(\widetilde
J_{\mu}(q) \, J_{\nu}(0))|P,  {\cal {Q}}\rangle \, ,
\end{equation}
where $J_\mu$ and $J_\nu$ are the electromagnetic current operators.
In addition, $P$ is the four-momentum of the initial hadronic state,
$q$ is the four-momentum of the virtual photon, and ${\cal {Q}}$ is
the charge of the hadron.
${\widehat{T}}(\widehat{{\cal{O}}}_{1}\widehat{{\cal{O}}}_{2})$
indicates time-ordered product between the operators
$\widehat{{\cal{O}}}_{1}$ and $\widehat{{\cal{O}}}_{2}$, and the
Fourier transform is indicated with a tilde. The tensor
$T_{\mu\nu}\equiv T_{\mu\nu}(P, q, h)$ has identical symmetry
properties as $W_{\mu\nu}(P, q, h)$, thus having similar
Lorentz-tensor structure to $W_{\mu\nu}$. By using the optical
theorem one obtains
\begin{equation}\label{DIS51}
\textmd{Im}\,\widetilde{F_{j}}=2\pi\,F_{j}\,,
\end{equation}
where $\widetilde{F_{j}}$ is the $j$-th structure function of the
$T_{\mu\nu}$ tensor, while $F_{j}$ is the one corresponding to the
$W_{\mu\nu}$ tensor.

%---------------------------------------------------------------------------
\section{DIS from scalar and vector mesons with $N_f=1$}\label{ScalarVector}
%---------------------------------------------------------------------------

%---------------------------------------------------------------------------
\subsection{General background}
%---------------------------------------------------------------------------

In this section we study a general approach to obtain the structure
functions for scalar and vector mesons with a single flavor,
$N_f=1$, in terms of two-point correlation functions of global
$U(1)$ symmetry currents. This is a holographic dual approach based
on \cite{Polchinski:2002jw}. In particular, we show that the
structure functions can be written as the product of a
model-dependent factor times a model-independent one. We explicitly
calculate both factors in terms of the parameters defining a general
holographic dual model. This includes the structure functions
derived from the D3D7-brane model and from the
D4D8$\mathrm{\overline{D8}}$-brane model that we already obtained in
our previous paper \cite{Koile:2011aa}, as well as those obtained
from the D4D6$\mathrm{\overline{D6}}$-brane model which we introduce
in appendix B of the present work.

Let us consider a general ten-dimensional background metric in the
Einstein frame written as
\begin{eqnarray}\label{zetas}
ds^{2} = \bigg(\frac{\rho}{R}\bigg)^\alpha \, \eta_{\mu\nu}dx^\mu
dx^\nu + \bigg(\frac{\rho}{R}\bigg)^\beta d\overrightarrow{Z}\cdot
d\overrightarrow{Z} \, ,
\end{eqnarray}
where $x^\mu=(x^0,\ldots,x^3)$, while
$\overrightarrow{Z}=(Z^1,\ldots,Z^6)$, with $\alpha>1$, $\beta<-1$.
We then add a probe Dp-brane with an induced metric of the form
\begin{eqnarray}\label{metgral}
ds^{2}_{Dp} = \bigg(\frac{\rho}{R}\bigg)^\alpha \, \eta_{\mu\nu} \,
dx^\mu dx^\nu + \bigg(\frac{\rho}{R}\bigg)^\beta \bigg[d\rho^2 +
\rho^2 \, d\Omega^2_{p-4}\bigg] \, ,
\end{eqnarray}
where $\rho$ is the radial direction of Dp-brane world-volume. The
radius $R$ is the length scale of the system, while $\Omega_{p-4}$
indicates coordinates on $S^{p-4}$.

This general induced metric also describes the D3D7,
D4D8$\mathrm{\overline{D8}}$, and D4D6$\mathrm{\overline{D6}}$-brane
models. In particular, for the D3D7-brane model we must set $p=7$,
$\alpha=2$, and $\beta=-2$. The asymptotic geometry is AdS$_5\times
S^5$, and $R$ gives the sphere and AdS$_5$ radii. In the case of the
D4D8$\mathrm{\overline{D8}}$-brane system we set $p=8$,
$\alpha=\frac{3}{2}$, and $\beta=-\frac{3}{2}$. In addition, for the
D4D6$\mathrm{\overline{D6}}$-brane model we have $p=6$,
$\alpha=\frac{3}{2}$, and $\beta=-\frac{3}{2}$. In all these cases,
we only recover the asymptotic metric, {\it i.e.} for $\rho \gg
\rho_0$ ($U\gg U_0$ in the notation of \cite{Sakai:2004cn}), which
is the relevant induced metric to the DIS process.

Scalar and vector mesons correspond to excitations of open strings
ending on the probe Dp-brane. The dynamics of the Dp-brane
fluctuations is described by the action
\begin{eqnarray}\label{kru4}
S_{Dp} &=& - \mu_{p} \int d^{p+1}\xi \, \sqrt{-\textmd{det}(\hat
P[g]_{ab} + 2 \pi \alpha'F_{ab})} + \frac{(2 \pi \alpha')^{2}}{2}
\mu_{p} \int \hat P[C^{(p-3)}] \wedge
F \wedge F \, , \nonumber \\
&&
\end{eqnarray}
where $g_{ab}$ stands for the metric (\ref{metgral}),
$\mu_{p}=\big[(2\pi)^{p}g_{s}\alpha'^{\frac{p+1}{2}}\big]^{-1}$ is
the Dp-brane tension and $\hat P$ denotes the pullback of the
background fields on the Dp-brane world-volume.

%--------------------------------------------------------------------
\subsection{DIS from scalar mesons}\label{Scalar}
%--------------------------------------------------------------------

The equations of motion for scalar mesons are obtained from
fluctuations of the probe Dp-brane which are orthogonal to the
directions of the brane world-volume. Let us take a coordinate $Z^i$
in Eq.(\ref{zetas}), which is perpendicular to the Dp-brane
world-volume, and slightly perturb it as follows
\begin{equation}\label{pert}
Z^{i}=Z^{i}_0 + 2 \pi \alpha' \Phi \, ,
\end{equation}
where $\Phi$ is a scalar fluctuation whose Lagrangian is
straightforwardly derived from the action of Eq.(\ref{kru4}), by
setting $F_{ab}=0$. By expanding to second order in the fluctuation,
one obtains
\begin{equation}
S^{scalar}_{0}=-\mu_{p} \int d^{p+1}\xi \, \sqrt{-\det g} \,
\bigg[1+\frac{(2\pi\alpha')^2}{2}\bigg(\frac{\rho}{R}\bigg)^\beta
g^{ab} \, \partial_a\Phi \, \partial_b\Phi\bigg] \, ,
\end{equation}
which corresponds to the Lagrangian
\begin{equation}\label{scafree}
{\mathcal{L}}^{scalar}_{0}=-\mu_{p} \, \sqrt{- \det g} \,
\bigg[1+\frac{(2\pi\alpha')^2}{2}\bigg(\frac{\rho}{R}\bigg)^\beta
g^{ab} \, \partial_a\Phi \, \partial_b\Phi\bigg] \, ,
\end{equation}
where all indices denote directions along the Dp-brane world-volume.
The probe brane wraps a $S^{p-4}$. By plugging the metric
(\ref{metgral}) into the quadratic Lagrangian, one obtains the
equations of motion (EOM) for scalar fluctuations of the Dp-brane in
the probe approximation
\begin{equation}\label{kru8}
\partial_{a}\bigg[\bigg(\frac{\rho}{R}\bigg)^{\theta-\beta}\sqrt{\widetilde{g}}
\, g^{ab} \, \partial_b\Phi\bigg] = 0 \, ,
\end{equation}
where we have defined
\be\label{theta}
\theta = 2 \, \alpha + \bigg(\frac{p}{2} - \frac{3}{2}\bigg) \beta +
(p-4) \, .
\ee
Notice that $\widetilde{g}_{ij}$ is the metric on $S^{p-4}$, which
together with $\rho$ span coordinates ($Z^{1}$, $\cdots$,
$Z^{p-3}$). The EOM can be more explicitly written as
\begin{eqnarray}\label{kru9}
\Box\Phi + \bigg(\frac{\rho}{R}\bigg)^{\alpha-\beta-2} R^{-2} \,
\nabla_i\nabla^i\Phi + \theta R^{-1} \,
\bigg(\frac{\rho}{R}\bigg)^{\alpha-\beta-1}\partial_\rho\Phi +
\bigg(\frac{\rho}{R}\bigg)^{\alpha-\beta}\partial_\rho^2\Phi=0 \, ,
\end{eqnarray}
where $\nabla_{i}$ is the covariant derivative on $S^{p-4}$.

We propose the following {\it Ansatz}\footnote{This solution will be
exact for the metric (\ref{metgral}). Since this metric is only
asymptotic for the models that we consider, {\it i.e.} D3D7,
D4D8$\mathrm{\overline{D8}}$, and D4D6$\mathrm{\overline{D6}}$-brane
models, we must assume the condition $\rho_{int}\sim q R^{2}\gg
\rho_{0} \equiv\Lambda R^{2}$, where $\rho_{int}$ denotes the
interaction region, while $\Lambda$ is an infrared cutoff of the
four-dimensional gauge theory.}
\begin{equation}\label{kru12}
\Phi^{\ell} = \phi^\ell(\rho) \, e^{i P \cdot y} \, Y^{\ell}(S^{p-4}) \,
,
\end{equation}
where $Y^{\ell}(S^{p-4})$ are the scalar spherical harmonics on
$S^{p-4}$, which satisfy the eigenvalue equation
\begin{equation}\label{kru13}
\nabla^{i}\nabla_{i}Y^{\ell}(S^{p-4}) = - \ell \, (\ell+p-5) \,
Y^{\ell}(S^{p-4}) \, .
\end{equation}
Now, by replacing the {\it Ansatz} (\ref{kru12}) in the EOM
(\ref{kru9}), we obtain
\begin{equation}\label{kru20}
\Phi_{IN/OUT}^\ell = c_{i} \, \bigg(\frac{\rho}{R}\bigg)^{A-\gamma
B} \, e^{i P \cdot y} \, Y^\ell(S^{p-4}) \, ,
\end{equation}
\begin{equation}\label{kru21}
\Phi_{X}^\ell = c_{X} \, s^{1/4} \, \Lambda^{-1/2} \,
\bigg(\frac{\rho}{R}\bigg)^{A} \,
J_\gamma\bigg[\frac{s^{1/2}R}{B}\bigg(\frac{\rho}{R}\bigg)^{-B}\bigg]
\, e^{i P \cdot y} \, Y^\ell(S^{p-4}) \, ,
\end{equation}
were we have used the full solution for $\Phi$ in the second case,
corresponding to the intermediate state $X$, and the leading
behavior in the region $\rho \sim \rho_{int}$ for the initial/final
hadronic state (IN/OUT). $J_{\gamma}$ is the Bessel function of
first kind, and $s=-(P+q)^2=M_X^2$ is the mass-squared of the
intermediate state, while $c_{X}$ and $c_{i}$ are dimensionless
constants. The order of the Bessel function is given by
\be
\label{nu} \gamma^2=\frac{A^2+\ell(\ell+p-5)}{B^2} \, ,
\ee
with the definitions
\be\label{defsAB}
A=\frac{1-\theta}{2}\quad ,  \quad B=\frac{\alpha-\beta-2}{2} \, .
\ee
These are scalar and pseudoscalar mesons for even and odd values of
$\ell$, respectively. This can be seen from the fact that under
parity transformation the spherical harmonics satisfy the equation
$[Y^\ell(S^{p-4})]_{\textmd{P}}=(-1)^{\ell} \, Y^\ell(S^{p-4})$.

By applying the method developed in \cite{Koile:2011aa}, we couple
these holographic scalar mesons to a gauge field in the bulk. This
is done by considering a metric fluctuation as given in
Eq.(\ref{pol5}). Then, we use the eigenvalue equation $\upsilon^{j}
\,
\partial_{j}Y^\ell(\Omega) = i \, {\cal{Q}}_\ell \, Y^\ell(\Omega)$,
obtaining the interaction Lagrangian\footnote{We have dropped the
label $\ell$ corresponding to the spherical harmonics from the
scalar field $\Phi^\ell$, as well as from the charges
${\cal{Q}}_\ell$.}
\begin{equation}\label{kru15}
{\mathcal{L}}_{interaction}^{scalar} = i \, {\cal{Q}} \, \mu_{p} \,
(\pi \alpha')^{2} \, \sqrt{-\det g} \,
\bigg(\frac{\rho}{R}\bigg)^{\beta} A^{m} \, (\Phi \,
\partial_{m}\Phi^{*}_{X} - \Phi^{*}_{X} \, \partial_{m}\Phi ) \, .
\end{equation}
The angular dependence on the spherical harmonics corresponds to
functions which are charge eigenstates, with charge ${\cal{Q}}$
under the $U(1)$ symmetry group which is induced by transformations
on the internal $S^{p-4}$ in the direction of the Killing vector
$\upsilon^{j}$.

Alternatively, as we explained in \cite{Koile:2011aa},
${\cal{L}}_{interaction}^{scalar}$ can be obtained from the coupling of the
gauge field $A_{m}$ to the Noether's current corresponding to the
global transformations which leave invariant the Lagrangian
(\ref{scafree}). These are transformations of a $U(1)\subseteq
SO(p-3)$, being the latter the isometry group of $S^{p-4}$. The
referred Noether's current is
\begin{equation}\label{kru18}
j_{m}^{scalar} = i \, \mu_{p} \, (\pi \alpha')^{2} \,
\bigg(\frac{\rho}{R}\bigg)^{\beta} \, (\Phi \,
\partial_m\Phi^{*}_{X} - \Phi^{*}_{X} \, \partial_{m}\Phi ) \, ,
\end{equation}
and by defining ${\cal{L}}_{interaction}^{scalar} = {\cal{Q}} \,
\sqrt{-\det g} \,
 A^{m} \, j_{m}^{scalar}$, we obtain the
same ${\cal{L}}_{interaction}^{scalar}$ given by Eq.(\ref{kru15})
from the metric fluctuation. Consequently,
${\cal{L}}_{interaction}^{scalar}$ is given by the coupling of the
gauge field $A^{m}$ to the conserved Noether's current
$j_{m}^{scalar}$. Notice that the scalar fields in Eq.(\ref{kru12})
are charged under the global $U(1)\subseteq SO(p-3)$, with
${\cal{Q}}_\ell \neq 0$ for $\ell>0$.

Now let us obtain the relevant matrix element for the hadronic
tensor, with the prescription proposed in \cite{Polchinski:2002jw}
\begin{equation}\label{kru21.1}
S_{interaction}=(2\pi)^4 \, \delta^4(P_X-P-q) \, \tilde{n}_{\mu} \, \langle
P+q, X|J^{\mu}(0)|P, {\cal {Q}}\rangle \, ,
\end{equation}
where $\tilde{n}_\mu$ indicates the polarization unit vector.

In order to calculate the gauge field we have to solve the Maxwell's
equation $D_m F^{mn}=0$, where $m, n = 0, 1, 2, 3, \rho$. We propose
the {\it Ans\"{a}tze}
\ba\label{gaugeansatz}
A_\mu&=&\widetilde{n}_\mu \, e^{i q \cdot y} \, f(\rho) \, , \nonumber \\
A_\rho&=& e^{i q \cdot y} \, g(\rho) \, ,
\ea
which imply a Lorentz-like gauge. The solution is
\begin{eqnarray}\label{gaugesol}
A_{\mu}&=& \widetilde{n}_{\mu} \, e^{i q \cdot y} \,
\frac{1}{\Gamma(n+1)} \, \bigg(\frac{qR}{2B}\bigg)^{n+1} \,
\bigg(\frac{\rho}{R}\bigg)^{-(n+1)B}
\, K_{n+1} \, \bigg[\frac{q R}{B} \bigg(\frac{\rho}{R} \bigg)^{-B} \bigg] \, , \nn\\
A_{\rho}&=&-e^{i q \cdot y} \, \frac{i(q\cdot
\widetilde{n})}{\Gamma(n+1)} \, \bigg(\frac{qR}{2B}\bigg)^{n+1} \,
\bigg(\frac{\rho}{R}\bigg)^{D} \, K_{n} \, \bigg[\frac{q
R}{B}\bigg(\frac{\rho}{R}\bigg)^{-B}\bigg] = - \frac{i}{q^{2}} \,
\eta^{\mu\nu} \, q_{\mu}\partial_{\rho} \, A_{\nu} \, ,
\end{eqnarray}
with
\be\label{Dn}
D = \frac{-4\alpha+3\beta+2}{4}\quad , \quad n=\frac{2+\beta}{4B} \,
,
\ee
and $B$ is given in Eq.(\ref{defsAB}). The current conservation
equation reads\footnote{Note that the subindex $\rho$ only indicates
the variable $\rho$, thus there is no sum whenever it appears
repeated.}
\be
\label{currentcons} \partial\cdot j^{scalar} =
\bigg(\frac{\rho}{R}\bigg)^{-\alpha-\frac{(p+3)}{2}\beta-(p-4)} \,
\partial_\rho\bigg[\bigg(\frac{\rho}{R}\bigg)^{2\alpha + \frac{(p-5)}{2} \beta
+ (p-4)}j^{scalar}_\rho\bigg] = 0 \, ,
\ee
while the coupling is
\begin{equation}\label{kru100}
A_{m} \, j_{scalar}^{m} = \bigg(\frac{\rho}{R}\bigg)^{-\alpha} \,
A_{\mu} \, \bigg[j^{\mu}_{scalar} - i \, \frac{q^{\mu}}{q^{2}} \,
(\partial\cdot j_{scalar})\bigg] - i \, \frac {q^{\nu}}{q^{2}} \,
\bigg(\frac{\rho}{R}\bigg)^{3-\theta} \,
\partial_{\rho} \bigg[\bigg(\frac{\rho}{R}\bigg)^{\theta-1}j_\rho^{scalar}
A_{\nu}\bigg]\,.
\end{equation}
Then, the interaction reads
\begin{eqnarray}\label{kru101}
S_{interaction}^{scalar} &=& {\mathcal{Q}} \int_{\rho_{0}}^{\infty} d^{p+1}x \,
\sqrt{-\textmd{det} g} \, A_{m} \, j^{m}_{scalar} \nonumber\\
&=& {\mathcal{Q}} \int_{\rho_{0}}^{\infty} d^{p+1}x \,
\sqrt{-\textmd{det} g} \, \bigg(\frac{\rho}{R}\bigg)^{-\alpha} \,
A_{\mu} \, \bigg[j_{scalar}^{\mu} - i
\frac{q^{\mu}}{q^{2}} \, (\partial\cdot j_{scalar})\bigg] + \nonumber \\
&& \frac {q^{\nu}}{q^{2}} \, {\mathcal{Q}} \int_{\rho_{0}}^{\infty}
d^{p+1}x \, \sqrt{-\textmd{det} g} \,
\bigg(\frac{\rho}{R}\bigg)^{-\theta}\partial_\rho\bigg[\bigg(\frac{\rho}{R}\bigg)^{\theta-1}j_\rho^{scalar}
\, A_{\nu}\bigg]
\, \nonumber \\
&\equiv&I_1^{scalar}+I_2^{scalar} \, .
\end{eqnarray}
It can be seen that in the limit $\Lambda\ll q$,
$I_2^{scalar}\rightarrow 0$. On the other hand, by evaluating
$I_1^{scalar}$ and using the Ansatz (\ref{kru21.1}), we find
\ba\label{kru102}
\langle P+q, X|J^{\mu}(0)|P, {\cal {Q}}\rangle \, &=& 2^{\gamma+2}
\,  B^{\gamma+1} \, \pi^2 \, \frac{\Gamma(\gamma+n+2)}{\Gamma(n+1)} \,  c_i \, c_X^* \, \mu_p
\, {\mathcal{Q}} \, \alpha'^2 \,  (s^{1/4}\Lambda^{-1/2})
\times \nonumber \\
&& \frac{q^{2n+2} \, s^{\frac{\gamma}{2}} \,
R^{p-\gamma-5}}{(q^2+s)^{2+\gamma+n}} \,
\bigg(P^\mu+\frac{q^\mu}{2x}\bigg) \, .
\ea
For $|t|\ll 1$ we can approximate $s \simeq q^{2}(1/x-1)$, thus the
above expression becomes
\ba\label{matrixelesc}
\langle P+q, X|J^{\mu}(0)|P, {\cal {Q}}\rangle & = & 2^{\gamma+2} \, B^{\gamma+1} \, \pi^2 \,
 \frac{\Gamma(\gamma+n+2)}{\Gamma(n+1)} \, c_i \, c_X^* \, \mu_p \,  {\mathcal{Q}} \,
 \alpha'^2 \,  R^{p-3} \times \nonumber\\
&&  \bigg(\frac{\Lambda}{q}\bigg)^{\gamma + \frac{3}{2}} \,
x^{\frac{\gamma}{2} + \frac{7}{4}+n}(1-x)^{\frac{\gamma}{2} +
\frac{1}{4}} \, \bigg(P^\mu+\frac{q^\mu}{2x}\bigg) \, .
\ea

Following \cite{Polchinski:2002jw} and \cite{Koile:2011aa}, we can
calculate $\textmd{Im}\:T^{\mu\nu}$ by multiplying
Eq.(\ref{matrixelesc}) by its complex conjugate and summing over
radial excitations. We estimate the density of states by introducing
an IR cutoff at $\rho_{0}\equiv \Lambda R^2$. The distance between
zeros of the Bessel function of Eq.(\ref{kru21}) is $M_{n'}=n' \pi
\Lambda$, which in the large $N$ limit and for large $q$ gives
\begin{eqnarray}\label{kru103}
\sum_{n'}\delta(M_{n'}^{2}-s)\sim \bigg(\frac{\partial
M_{n'}^{2}}{\partial n'}\bigg)^{-1}\sim (2\pi s^{1/2}\Lambda)^{-1}
\, .
\end{eqnarray}
Finally, we obtain
\begin{eqnarray}
\label{kru22}
\textmd{Im} \:T^{\mu\nu} &=&  2^{2\gamma+4} \, B^{2\gamma+2} \, \pi^5 \,
 \frac{\Gamma(\gamma+n+2)^2}{\Gamma(n+1)^2} \, |c_i|^2 \,|c_X|^2 \, \mu_p^2 \,
 {\mathcal{Q}}^2 \, \alpha'^4 \, R^{2p-4}  \times \\
&& \bigg(\frac{\Lambda^2}{q^2}\bigg)^{\gamma+2} \, x^{\gamma+4+2n} \, (1-x)^{\gamma}
\, \bigg(P^\mu+\frac{q^\mu}{2x}\bigg) \, \bigg(P^\nu+\frac{q^\nu}{2x}\bigg)
\,. \,\,\,\,\,\,\,\,\: \,\,\,\,\,\,\,\,\:\nonumber
\end{eqnarray}
After checking that our $W^{\mu\nu}$ satisfies all the symmetry
requirements described above, we obtain the structure functions for
the scalar mesons from Eq.(\ref{DIS51}):
\begin{equation}\label{kru23}
F_{1}=0, \,\,\,\,\,\,\,\,\:\,\,\,\,\,\,\,\,\:\:\:\:F_{2}=
A_{0}^{scalar}\mu_p^2{\mathcal{Q}}^2\alpha'^4R^{2p-6}\bigg(\frac{\Lambda^2}{q^2}\bigg)^{\gamma+1}
x^{\gamma+3+2n}(1-x)^{\gamma} \, ,
\end{equation}
where $A_{0}^{scalar}= 2^{2\gamma+4} \, B^{2\gamma+2} \, \pi^5 \,
\frac{\Gamma(\gamma+n+2)^2}{\Gamma(n+1)^2} \, |c_i|^2 \, |c_X|^2$ is
a dimensionless normalization constant. We can easily check that our
previous results for D3D7 and D4D8$\mathrm{\overline{D8}}$-brane
systems introduced in \cite{Koile:2011aa} are recovered. Also, for
the D4D6$\mathrm{\overline{D6}}$-brane system we obtain the results
shown in appendix B of the present work.

%--------------------------------------------------------------------
\subsection{DIS from vector mesons}\label{Vector}
%--------------------------------------------------------------------

In this subsection we calculate the hadronic tensor for vector
mesons arising from a single probe brane, {\it i.e.} $N_f=1$. Next,
we will decompose this tensor in order to obtain the structure
functions. The procedure will be analogous to that developed in last
subsection, though the calculations are more tedious.

Vector mesons arise from fluctuations of the vector fields on the
Dirac-Born-Infeld (DBI) action of the probe Dp-brane, which are in
the directions parallel to the brane world-volume
\cite{Kruczenski:2003be}. The starting point is the action
(\ref{kru4}). We calculate the EOM for vector fluctuations, keeping
$Z^i$ constant, \emph{i.e.} $\Phi=0$, and then by expanding the
Lagrangian up to quadratic order in the fluctuation. This new
Lagrangian gives the following EOM
\begin{eqnarray}\label{kru26}
\partial_{a}\bigg(\sqrt{-\textmd{det} g}F^{ab}\bigg)=0 \, ,
\end{eqnarray}
where $F^{ab}=\partial^{a}B^{b} - \partial^{b}B^{a}$, and the
indices $a, b = 0, \ldots, p$ run over all directions within the
Dp-brane world-volume. We have considered only the DBI term of
Eq.(\ref{kru4}), since the Wess-Zumino term does not contribute at
first order for the set of solutions in which we are interested. By
expanding Eq.(\ref{kru26}) we can write
\begin{equation}\label{vectexpand}
\Box B^\mu-\partial^\mu(\partial\cdot B)+\theta
R^{-1}\bigg(\frac{\rho}{R}\bigg)^{\alpha-\beta-1}\partial_\rho
B^\mu+
\bigg(\frac{\rho}{R}\bigg)^{\alpha-\beta}\partial_\rho^2B^\mu+\nabla_i\nabla^iB^\mu=0.
\end{equation}
We propose the same Ansatz used in \cite{Kruczenski:2003be} for the
solution of vector mesons $B_{\mu}$
\begin{eqnarray}\label{vectAns}
B_{\mu}^\ell=\zeta_{\mu}\:\phi^\ell(\rho)\:e^{iP \cdot
y}\:Y^{\ell}(S^{p-4}),\:\:\:\:\:P \cdot
\zeta=0,\:\:\:\:\:B_{\rho}=0,\:\:\:\:\:B_{i}=0 \, ,
\end{eqnarray}
where it has been done an expansion in $Y^{\ell}(S^{p-4})$, which
are spherical harmonics on $S^{p-4}$ satisfying Eq.(\ref{kru13}).
$\phi^\ell(\rho)$ is a function to be determined, $\zeta_{\mu}$ is
the polarization vector and the relation $\zeta \cdot P=0$ comes
from $\partial^{\mu}B_{\mu}=0$.

By plugging the {\it Ansatz} (\ref{vectAns}) in
Eq.(\ref{vectexpand}), we obtain
\begin{equation}\label{solvectin}
B_{\mu IN/OUT}^\ell=\zeta_\mu
\Lambda^{-1}c_{i}\bigg(\frac{\rho}{R}\bigg)^{A-\gamma B}e^{iP \cdot
y} \, Y^\ell(S^{p-4}) \, ,
\end{equation}
\begin{equation}\label{solvectx}
B_{\mu X}^\ell=\zeta_{\mu X}
\Lambda^{-1}c_{X}(s^{-1/4}\Lambda^{1/2})\bigg(\frac{\rho}{R}\bigg)^{A}
J_\gamma\bigg[\frac{s^{1/2}R}{B}\bigg(\frac{\rho}{R}\bigg)^{-B}\bigg]e^{iP
\cdot y} \, Y^\ell(S^{p-4})\,.
\end{equation}
We have used the full solution for $B_\mu^\ell$ in the second case,
corresponding to the intermediate state $X$ and the leading behavior
in the region $\rho \sim \rho_{int}$ for the initial/final hadronic
state. As before, $J_{\gamma}$ is the Bessel function of first kind,
$\gamma^2=\frac{A^2+\ell(\ell+p-5)}{B^2}$ and $s=-(P+q)^2=M_X^2$ is
the mass-squared of the intermediate state, while $c_{X}$ and
$c_{i}$ are dimensionless constants. We have also used the
definitions for $\theta$, $A$, and $B$ in Eqs.(\ref{theta}) and
(\ref{defsAB}).  We can classify the solutions as vector mesons for
even values of $\ell$, and axial vector mesons for odd values of
$\ell$. This comes from the relations
$[Y^{\ell}(\Omega)]_{\textmd{P}}=(-1)^{\ell} \, Y^{\ell}(\Omega)$
and $[\zeta_{\mu}]_{\textmd{P}}=(\zeta^{0}, -\vec{\zeta})$.

From the expansion of $B_{\mu}$ in spherical harmonics on $S^{p-4}$
it can be seen that the gauge fields on the Dp-brane correspond to
charged fields in ${\mathcal{M}}_5$. Following an analogous
procedure as in \cite{Koile:2011aa} it can be seen that the modes
with $\ell=0$ correspond to an Abelian gauge field $B_\mu^0$. The
rest of the vector fields, $B_\mu^\ell$ with $\ell>0$, are charged
massive fields. Their charges under the $U(1)\subseteq SO(p-3)$ of
$S^{p-4}$ are ${\cal{Q}}_\ell$, while their masses are
$m^{2}_{\ell}=\ell(\ell+p-5)/R^2$. The EOM for the vector mesons in
the interaction region, Eq.(\ref{vectexpand}), can also be derived
from the following quadratic Lagrangian,\footnote{We use
$B_{\mu}^{\ell}\equiv B_{\mu}$, with $\ell>0$, therefore there is a
field $B_{\mu}$ for each $\ell$. The superscript $SF$ stands for
single-flavored ($N_f=1$) vector mesons.}
\begin{eqnarray}\label{kru41}
{\mathcal{L}}_{0}^{SF}=-\mu_{p} \, (\pi \alpha')^{2} \, \sqrt{-\det
g} \, \, F^{ab} \, F_{ab}^{*}\,.
\end{eqnarray}
We reproduce the bulk interaction as we have done in the last
subsection, by perturbing the metric with the fluctuation
(\ref{pol5}), as explained before. We use again $\upsilon^{j} \,
\partial_{j}Y^\ell(\Omega) = i \, {\cal{Q}}_\ell \, Y^\ell(\Omega)$,
obtaining the interaction Lagrangian
\begin{equation}\label{kru42}
{\mathcal{L}}_{interaction}^{SF}=i \, {\cal{Q}} \, \mu_{p} \, (\pi
\alpha')^{2} \, \sqrt{-\det g} \,  A_{m} \, [B_{Xn}^{*} \, F^{nm} -
B_{n} \, (F_{X}^{nm})^{*}] \, ,
\end{equation}
where $A_m$ is the five-dimensional gauge field given in
Eq.(\ref{gaugesol}). As in last subsection, the same interaction
Lagrangian can be obtained from the coupling of the gauge field
$A_{m}$ to the Noether's current corresponding to the internal
global symmetry of the action, in this case Eq.(\ref{kru41}). We can
write ${\cal{L}}_{interaction}^{SF} = {\cal{Q}} \,
\sqrt{-\textmd{det} g} \, A_{m} \, j_{SF}^{m}$ where
\begin{equation}\label{kru43}
j_{SF}^{m} = i \, \mu_{p} \, (\pi \alpha')^{2} \, [B_{Xn}^{*} \,
F^{nm} - B_{n} \, (F^{nm}_{X})^{*}] \, ,
\end{equation}
is a conserved current. Following a similar procedure as for scalar
mesons, we have the action of the interaction
\begin{eqnarray}\label{kru101}
S_{interaction}^{SF} &=& {\mathcal{Q}} \, \int_{\rho_{0}}^{\infty} d^{p+1}x \,
\sqrt{-\textmd{det} g} \, A_{m} \, j_{SF}^{m} \nonumber\\
&=& {\mathcal{Q}} \, \int_{\rho_{0}}^{\infty} d^{p+1}x \,
\sqrt{-\textmd{det} g} \, \bigg(\frac{\rho}{R} \bigg)^{-\alpha} \,
A_{\mu}\bigg[j_{SF}^{\mu} - i \,
\frac{q^{\mu}}{q^{2}} \, (\partial\cdot j_{SF}) \bigg] + \nonumber \\
&& \frac {q^{\nu}}{q^{2}} \, {\mathcal{Q}} \,
\int_{\rho_{0}}^{\infty} d^{p+1}x \, \sqrt{-\textmd{det} g} \,
\bigg(\frac{\rho}{R} \bigg)^{-\theta} \,
\partial_\rho \bigg[\bigg(\frac{\rho}{R} \bigg)^{\theta-1} \, j_{\rho}^{SF}
\, A_{\nu} \bigg] \, \\
&\equiv& I_{1}^{SF}+I_{2}^{SF} \, .
\end{eqnarray}
In particular, $I_2^{SF}\rightarrow 0$ when $\Lambda\ll q$. By
evaluating $I_{1}^{SF}$ and using the Ansatz (\ref{kru21.1}), we
find
\ba
\label{kru102}
\langle P+q, X|J^{\mu}(0)|P, {\cal {Q}}\rangle \, &=& 2^{\gamma} \, B^{\gamma+1}
\ \pi^2 \frac{\Gamma(\gamma+n+2)}{\Gamma(n+1)} \, c_i \, c_X^* \,
\mu_p \,  {\mathcal{Q}} \, \alpha'^2 \, R^{p-\gamma-5} \times \nonumber \\
&&
 (s^{-1/4} \Lambda^{1/2})
\,  \frac{q^{2\gamma +2} s^{\frac{\gamma}{2}}}{(q^2+s)^{2+\gamma+n}} \, N^\mu \, ,
\ea
with
\begin{equation}\label{kru46a}
N^{\mu}=2 \, (\zeta \cdot
\zeta_{X})\bigg(P^{\mu}+\frac{q^{\mu}}{2x}\bigg)+(\zeta_{X} \cdot
q)\zeta^{\mu}-(\zeta \cdot q)\zeta^{\mu}_{X} \, .
\end{equation}
We can see that $N^\mu$, which carries all the information about the
vector dependence in the matrix element (\ref{kru102}), and
therefore in the structure functions, does not depend on the
particular model.

For $|t|\ll 1$ we can approximate $s \simeq q^{2}(\frac{1}{x}-1)$,
and we obtain
\begin{eqnarray}\label{matrixelvect}
\langle P+q, X|J^{\mu}(0)|P, {\cal {Q}}\rangle \,& = & 2^{\gamma} \, B^{\gamma+1} \, \pi^2 \,
\frac{\Gamma(\gamma+n+2)}{\Gamma(n+1)} \, c_i \, c_X^* \,
\mu_p \, {\mathcal{Q}} \, \alpha'^2 \, R^{p-3} \, \times
\nonumber \\
& & \bigg(\frac{\Lambda}{q}\bigg)^{\gamma+\frac{5}{2}} \, x^{\frac{\gamma}{2}
+\frac{9}{4}+n}(1-x)^{\frac{\gamma}{2}-\frac{1}{4}}N^\mu  \nonumber\\
& \equiv & f^{(\gamma)}_{\Lambda}(x,q) N^\mu.
\end{eqnarray}
We now multiply Eq.(\ref{matrixelvect}) by its complex conjugate and
sum over the radial excitations and over the polarizations of the
final hadronic states $\zeta_{X}^{\mu}$, since we want to calculate
$\textmd{Im}\:T^{\mu\nu}$ from Eq.(\ref{pol3}). The density of
states is estimated in the same way as we have done for the scalar
mesons, obtaining
\begin{eqnarray}\label{kru47}
\textmd{Im} \:T^{\mu\nu} &=& \frac{\pi f f^*}{\Lambda s^{1/2}} \,
\sum_{\lambda} \, N^{\mu} \, N^{*\nu} \, .
\end{eqnarray}
By using the solution (\ref{vectAns}), then we normalize the
polarizations as $\zeta^{\mu}(P_{X},\lambda) \cdot
\zeta_{\mu}^{*}(P_{X},\lambda')=-M^{2}_{X}\delta_{\lambda,\lambda'}$,
and by neglecting terms proportional to $q_{\mu}$ and $q_{\nu}$, we
finally obtain
\begin{eqnarray}\label{kru49}
\textmd{Im} T_{\mu\nu}&\equiv& 2\frac{\pi x^{\frac{1}{2}}f
f^*}{\Lambda q (1-x)^{\frac{1}{2}}} H_{\mu\nu} =2 \, \frac{\pi
x^{\frac{1}{2}}f f^*}{\Lambda q
(1-x)^{\frac{1}{2}}}(H_{\mu\nu}^{S}+H_{\mu\nu}^{A}) \, ,
\end{eqnarray}
where $H_{\mu\nu}^{S}$ and $H_{\mu\nu}^{A}$ are the symmetric and
antisymmetric parts of $H_{\mu\nu}$, respectively,
\begin{eqnarray}\label{kru50}
H_{\mu\nu}^{S}
&=&-\eta_{\mu\nu}(\zeta.q)(\zeta^{*}.q)(P+q)^{2}
+P_{\mu}P_{\nu}\bigg[-4P^{2}(P+q)^{2}+(q.\zeta)(q.\zeta^{*})\bigg]
\nonumber\\ & & +(\zeta_{\mu}\zeta_{\nu}^{*}+
\zeta_{\nu}\zeta_{\mu}^{*})\frac{1}{2}\bigg[(P\cdot q)^2-P^2
q^2\bigg] \nonumber\\& & +(P_{\mu}\zeta^{*}_{\nu}+P_{\nu}
\zeta^{*}_{\mu})(\zeta\cdot q)\frac{1}{2}\bigg[P \cdot q+q^{2}\bigg]
+(P_\mu \zeta_{\nu}+P_{\nu}\zeta_{\mu})(\zeta^{*} \cdot
q)\frac{1}{2}\bigg[P \cdot q+q^{2} \bigg],
\end{eqnarray}
and
\begin{eqnarray}\label{kru51}
H_{\mu\nu}^{A}&=&
\frac{1}{2}(\zeta_{\mu}\zeta_{\nu}^{*}-\zeta_{\nu}\zeta_{\mu}^{*})\bigg[(P\cdot
q)^2-P^2q^2\bigg]\nonumber\\ & &
+ \frac{1}{2} (P_{\nu}\zeta^{*}_{\mu} - P_{\mu} \zeta_{\nu}^{*})
(\zeta \cdot q) \bigg[4P^{2}+7P \cdot q+3q^{2}\bigg]\nonumber\\
& & + \frac{1}{2} (P_{\mu}\zeta_{\nu} - P_{\nu}
\zeta_{\mu})(\zeta^{*} \cdot q) \bigg[4 P^{2} + 7 P \cdot q + 3
q^{2} \bigg] \, .
\end{eqnarray}
It is straightforward to calculate the tensor $W_{\mu\nu}$ from
$\textmd{Im} T_{\mu\nu}$. By comparing the $W_{\mu\nu}$ tensor
obtained in this way with the general form of Eq.(\ref{DIS16}) we
can extract the eight structure functions (recall that we have
derived these equations for $|t|<<1$)
\begin{eqnarray}\label{strfunctabel}
F_{1}&=& A^{SF}(x)\frac{1}{12x^{3}}(1-x-2xt-4x^{2}t+4x^{3}t+8x^{3}t^{2})\,,\nn\\
F_{2}&=& A^{SF}(x)\frac{1}{6x^{3}}(1-x+12xt-14x^{2}t-12x^{2}t^{2})\,,\nn\\
b_{1}&=& A^{SF}(x)\frac{1}{4x^{3}}(1-x-xt)\,,\nn\\
b_{2}&=& A^{SF}(x)\frac{1}{2x^{3}}(1-x-x^{2}t)\,,\nn\\
b_{3}&=& A^{SF}(x)\frac{1}{24x^{3}}(1-4x+8x^{2}t)\,,\\
b_{4}&=& A^{SF}(x)\frac{1}{12x^{3}}(-1+4x-2x^{2}t)\,,\nn\\
g_{1}&=& A^{SF}(x)\frac{t}{8x^2}(-7+6x+8xt)\,,\nn\\
g_{2}&=& A^{SF}(x)\frac{1}{16x^{4}}(3-3x-4xt+2x^{2}t)\,,\nn
\end{eqnarray}
where
\begin{equation}
A^{SF}(x) =
A^{SF}_0\mu_p^2{\mathcal{Q}}^2\alpha'^4R^{2p-6}\bigg(\frac{\Lambda^2}{q^2}\bigg)^{\gamma}
x^{\gamma+2n+5}(1-x)^{\gamma-1}\,,
\end{equation}
and $A^{SF}_0=(2B)^{2\gamma+2} \, \pi^5 \,
\frac{\Gamma(\gamma+n+2)^2}{\Gamma(n+1)^2} \, |c_i|^2 \, |c_X|^2$ is
a dimensionless normalization constant. Recall that $\gamma$ is given in Eq.(\ref{nu}) and $n$ in
Eq.(\ref{Dn}). Constants $\alpha$ and $\beta$ come from the
definition of the general background metric (\ref{zetas}), $A$ and
$B$ are defined in Eqs.(\ref{defsAB}) and $\theta$ is given by
Eq.(\ref{theta}).

We can see that, as it happened with the scalar mesons, the results
for D3D7 and D4D8$\mathrm{\overline{D8}}$-brane systems from
\cite{Koile:2011aa} are recovered\footnote{We have found a mistake
in the equation for $g_1$ in our previous work \cite{Koile:2011aa}.
The present solution is the correct one.}, as well as that for
D4D6$\mathrm{\overline{D6}}$-brane system given in appendix B.

%---------------------------------------------------------------------------------
\section{DIS from vector mesons with $N_f>1$}\label{Non-Abelian}
%---------------------------------------------------------------------------------

%---------------------------------------------------------------------------------
\subsection{General background calculations}
%---------------------------------------------------------------------------------

In this section, we study a general approach to obtain the structure
functions for polarized vector mesons with $N_f>1$ flavors. From the
string theory dual model these mesons arise by considering $N_f>1$
probe Dp-branes. In particular, we show that the structure functions
can be decomposed in model-dependent and model-independent factors,
as it occurs when $N_f=1$. We calculate both factors for a general
model with an induced metric given by Eq. (\ref{metgral}). All the
calculations in this section are within the tree-level
approximation. One-loop corrections are discussed in section
\ref{LargeN}.

We consider the same background as in section \ref{ScalarVector},
given by the induced metric (\ref{metgral}) on the $N_f$ probe
Dp-branes
\begin{eqnarray}
ds^{2}=\bigg(\frac{\rho}{R}\bigg)^\alpha \, \eta_{\mu\nu}dx^\mu
dx^\nu + \bigg(\frac{\rho}{R}\bigg)^\beta \bigg[d\rho^2 + \rho^2 \,
d\Omega^2_{p-4}\bigg]\,,\nonumber
\end{eqnarray}
at least in the asymptotic region, $\rho_{int}\gg \rho_0= \Lambda
R^2$.

We start from the non-Abelian Dirac-Born-Infeld action
\cite{Tseytlin:1997csa}
\begin{eqnarray}\label{na1}
S_{0}^{MF} = - \mu_p \, \int \, d^{p+1}\xi \, \sqrt{-\textmd{det}g}
\, (\pi\alpha')^2 \, \textmd{Tr}(F^2) \, ,
\end{eqnarray}
where $F_{ab}=\partial_a B_b-\partial_{b} B_a+i \, [B_a,B_b]$ and
$\mu_p=[(2\pi)^p g_s \alpha'^{\frac{p+1}{2}}]^{-1}$. This is the
generalization of Eq.(\ref{kru41}) for the case of mesons with
$N_f>1$.

In order to calculate the hadronic tensor using the holographic dual
prescription we consider that the holographic meson couples to a
gauge field (\ref{gaugesol}) in the bulk of the string theory dual
model as in section \ref{ScalarVector}.

We can expand the action Eq.(\ref{na1}) in terms of $B_\mu$
obtaining\footnote{We write $\textmd{Tr}(F_{ab}^*F^{ab})$ instead of
$\textmd{Tr}(F_{ab} F^{ab})$ since their EOM's are the same.}
\begin{eqnarray}\label{freenonabelian}
{\mathcal{L}}_{0}^{MF} = - \mu_p \, (\pi\alpha')^2 \, \sqrt{-\det g}
\, Tr\bigg\{ \widehat{F}^*_{ab}\widehat{F}^{ab} + \bigg(\, i \,
\widehat{F}_{ab}^*[B^a,B^b]+c.c.\bigg)-[B_a^*,B_b^*][B^a,B^b]\bigg\},
\end{eqnarray}
where we have defined $\widehat{F}_{ab}=\partial_a B_b-\partial_{b}
B_a=F_{ab}- i \, [B_a,B_b]$. As we shall see in section
\ref{multipleflavormesons}, the last two terms are sub-leading with
respect to the first one in the $1/N$ expansion. Therefore, at
leading order we only keep the first term. Thus, we obtain the same
interaction Lagrangian as for vector mesons with $N_f=1$.

The EOM can be expanded and we obtain
\begin{equation}\label{naexpand}
\Box B^\mu-\partial^\mu(\partial\cdot B)+\theta \,
R^{-1}\bigg(\frac{\rho}{R}\bigg)^{\alpha-\beta-1}\partial_\rho
B^\mu+
\bigg(\frac{\rho}{R}\bigg)^{\alpha-\beta}\partial_\rho^2B^\mu+\nabla_i\nabla^iB^\mu=0,
\end{equation}
which is the same as Eq.(\ref{vectexpand}). The {\it Ansatz} for the
solution of the vector mesons $B_{\mu}$ is
\begin{eqnarray}\label{ansna}
B_{\mu}^\ell & = & \sum_{{\mathcal{A}}=1}^{N_f} B_\mu^{({\mathcal{A}})\ell} \, \tau_{\mathcal{A}} \, , \nonumber \\
B_{\mu}^{({\mathcal{A}})\ell}&=&\zeta_{\mu}\:c^{({\mathcal{A}})}\phi^\ell(\rho)\:e^{iP
\cdot y}\:Y^{\ell}(S^{(p-4)}),\:\:\:\:\:P \cdot
\zeta=0,\:\:\:\:\:B_{\rho}^{({\mathcal{A}})\ell}=0,\:\:\:\:\:B_{i}^{({\mathcal{A}})\ell}=0\,,
\end{eqnarray}
where $\tau_{\mathcal{A}}$ are the generators of the flavor group $SU(N_f)$,
which satisfy the Lie algebra
\be
[\tau_{\mathcal{A}},\tau_{\mathcal{B}}]=i \, f_{{\mathcal{ABC}}} \, \tau_{\mathcal{C}} \, .
\ee
We have also expanded $B^{({\mathcal{A}})}_\mu$ in spherical
harmonics $Y^{\ell}(S^{(p-4)})$, satisfying Eq.(\ref{kru13}). The
radial dependence $\phi(\rho)$ is to be determined, $\zeta_{\mu}$ is
the polarization vector and the relation $\zeta \cdot P=0$ comes
from $\partial^{\mu}B_{\mu}=0$. By using the {\it Ansatz}
(\ref{ansna}) in Eq.(\ref{naexpand}), we obtain the solution for
each component $B_\mu^{({\mathcal{A}})}$ which coincides with the
vector mesons with $N_f=1$ studied in previous section, namely
\begin{equation}\label{solvectinMF}
B_{\mu IN/OUT}^{({\mathcal{A}})\ell}=\zeta_\mu
\Lambda^{-1}c_{i}^{({\mathcal{A}})}\bigg(\frac{\rho}{R}\bigg)^{A -
\gamma B} \, e^{i P \cdot y} \, Y^\ell(S^{(p-4)}) \, ,
\end{equation}
\begin{equation}\label{solvectxMF}
B_{\mu X}^{({\mathcal{A}})\ell}=\zeta_{\mu X}
\Lambda^{-1}c_{X}^{({\mathcal{A}})}(s^{-1/4}\Lambda^{1/2})\bigg(\frac{\rho}{R}\bigg)^{A}
J_\gamma\bigg[\frac{s^{1/2}R}{B}\bigg(\frac{\rho}{R}\bigg)^{-B}\bigg]
\, e^{iP \cdot y} \, Y^\ell(S^{(p-4)})\,.
\end{equation}
We have used the full solution for $B_\mu^{({\mathcal{A}})\ell}$ in
the second case, corresponding to the intermediate state $X$ and the
leading behaviour in the region $\rho\sim\rho_{int}$ for the
initial/final hadronic state (IN/OUT). As before, $J_{\gamma}$ is
the Bessel function of first kind,
$\gamma^2=\frac{A^2+\ell(\ell+p-5)}{B^2}$ and $s=-(P+q)^2=M_X^2$ is
the mass-squared of the intermediate state, while
$c_{X}^{({\mathcal{A}})}$ and $c_{i}^{({\mathcal{A}})}$ are
dimensionless constants. We have also used the definitions for
$\theta$, $A$, and $B$ in Eqs. (\ref{theta}) and (\ref{defsAB}).
From the expansion of $B_{\mu}^{({\mathcal{A}})}$ in spherical
harmonics on $S^{(p-4)}$, it can be seen that the gauge fields on
the branes correspond to charged massive fields in the
five-dimensional space spanned by coordinates $0,1,2,3,$ and $\rho$,
for $\ell>1$, and a gauge field $B^0_\mu$.

By considering the metric fluctuation from Eq.(\ref{pol5}), and
equation $\upsilon^{j} \,
\partial_{j}Y^\ell(\Omega)=i \, {\cal{Q}}_\ell \, Y^\ell(\Omega)$, we
obtain the interaction Lagrangian\footnote{In what follows we denote
$B_\mu \equiv B_\mu^{\ell}$, {\it i.e.} we omit the $\ell$ index to
make our notation simpler. The label $MF$ stands for multi-flavored
vector mesons, {\it i.e.} those with $N_f>1$.}
\begin{eqnarray}\label{kru42bis}
{\mathcal{L}}_{interaction}^{MF}&=&i{\cal{Q}}\mu_{p}(\pi
\alpha')^{2} \, \sqrt{-\det g}
 \, \bigg\{A_{m}Tr\big(B_{Xn}^{*}\widehat{F}^{nm}-B_{n}(\widehat{F}_{X}^{nm}\big)^{*})+\nonumber\\
&&iA_{m}Tr\big(B_n[B_X^{m*},B_X^{n*}]\big)+iA_{m}Tr\big(B_{Xn}^*[B^{m},B^{n}]\big) \bigg\}\,\\
&\equiv
&{\mathcal{L}}_{interaction_1}^{MF}+{\mathcal{L}}_{interaction_2}^{MF}+{\mathcal{L}}_{interaction_3}^{MF}.\nonumber
\end{eqnarray}
It is easy to see that the term ${\mathcal{L}}_{interaction_3}^{MF}$
does not contribute to the process of interest, since it involves
two initial states for the hadron $B_\mu$. On the other hand, we
will show in the next section that the term
${\mathcal{L}}_{interaction_2}^{MF}$ contributes only to diagrams
which are sub-leading in the $1/N$ expansion. Therefore, the only
diagram which contributes to leading order is that of figure 1,
which only involves the first term
${\mathcal{L}}_{interaction_1}^{MF}$. This is the same diagram
present in the $N_f=1$ vector mesons studied in last section. We can
see it as the coupling of the gauge field $A_{m}$ to a certain
current $j_{MF}^m$. Therefore,
${\cal{L}}_{interaction_1}^{MF}={\cal{Q}} \, \sqrt{-\textmd{det}g}
\, A_{m} \, j_{MF}^{m}$ where
\begin{equation}\label{kru43}
j_{MF}^{m}=i \, \mu_{p} \, (\pi \alpha')^{2} \,
Tr\big(B_{Xn}^{*}F^{nm}-B_{n}(F^{nm}_{X})^{*}\big)\,.
\end{equation}
The action of interaction is then
\begin{eqnarray}\label{kru101}
S_{interaction}^{MF} &=& {\mathcal{Q}} \, \int_{\rho_{0}}^{\infty} d^{p+1}x
\, \sqrt{-\textmd{det} g} \, A_{m} \, j_{MF}^{m} \nonumber\\
&=& {\mathcal{Q}} \, \int_{\rho_{0}}^{\infty} d^{p+1}x \,
\sqrt{-\textmd{det} g} \, \bigg(\frac{\rho}{R}\bigg)^{-\alpha} \,
A_{\mu} \, \bigg[j_{MF}^{\mu} - i \, \frac{q^{\mu}}{q^{2}} \,
(\partial\cdot j_{MF})\bigg]+ \nonumber \\
&& \frac {q^{\nu}}{q^{2}}{\mathcal{Q}} \, \int_{\rho_{0}}^{\infty}
d^{p+1}x \, \sqrt{-\textmd{det}g} \,
\bigg(\frac{\rho}{R}\bigg)^{-\theta} \,
\partial_\rho\bigg[\bigg(\frac{\rho}{R}\bigg)^{\theta-1}j^{MF}_{\rho}
A_{\nu}\bigg]
\, \nonumber \\
&\equiv&I_{1}^{MF}+I_{2}^{MF} \, .
\end{eqnarray}
As we have seen, $I_{2}^{MF}=I_{2}^{SF}\rightarrow 0$ in the limit
of interest, $\Lambda\ll q$. On the other hand, by evaluating
$I_{1}^{MF}$ we can see that
\begin{eqnarray}\label{kru102na}
\langle P+q, X|J^{\mu}(0)|P, {\cal {Q}}\rangle \,&=& C_f \, \delta_{{\mathcal{AB}}} \, I_{1}^{SF}\nonumber\\
& = & C_f \, \delta_{{\mathcal{AB}}} \, (2B)^{\gamma+2} \, \pi^2 \, \frac{\Gamma(\gamma+n+2)}{\Gamma(n+1)}
\, c_i \, c_X^* \, \mu_p \, {\mathcal{Q}} \, \alpha'^2 \,  R^{p-\gamma-5} \nn\\
& &  (s^{-1/4}\Lambda^{1/2}) \, \frac{q^{2\gamma+2}s^{\frac{\gamma}{2}-n}}{(q^2+s)^{2+\gamma+n}} \, N^\mu \nn\\
&=& C_f \, \delta_{{\mathcal{AB}}} \, f^{(\gamma)}_{\Lambda}(x,q) \, N^\mu\,,
\end{eqnarray}
where we define $f^{(\gamma)}_{\Lambda}(x,q)$ as in last section and have used
\ba
B_{\mu IN/OUT}&=&B_{\mu IN/OUT}^{({\mathcal{A}})} \, \tau_{\mathcal{A}},\qquad \textmd{(no sum)}\nn\\
B_{\mu X}&=&B_{\mu X}^{({\mathcal{B}})} \, \tau_{\mathcal{B}}, \qquad \qquad \textmd{(no
sum)}
\ea
and
\begin{equation}\label{kru46a}
N^{\mu}=2(\zeta \cdot
\zeta_{X})\bigg(P^{\mu}+\frac{q^{\mu}}{2x}\bigg)+(\zeta_{X} \cdot
q)\zeta^{\mu}-(\zeta \cdot q)\zeta^{\mu}_{X}\,.
\end{equation}
We also use
\be
[\tau_{\mathcal{A}},\tau_{\mathcal{B}}]=i \, f_{{\mathcal{ABC}}} \, \tau_{\mathcal{C}}\ \quad , \quad
Tr(\tau_{\mathcal{A}}\tau_{\mathcal{B}})=C_f \, \delta_{{\mathcal{AB}}} \, ,
\ee
being $C_f$ the Casimir of $SU(N_f)$.

For $|t|\ll 1$ we can approximate $s\simeq q^{2}(\frac{1}{x}-1)$, as
we have done in the previous section
\begin{eqnarray}\label{matrixelnonab}
\langle P+q, X|J^{\mu}(0)|P, {\cal {Q}}\rangle \,&=& C_f \delta_{AB}{\mathcal{Q}}
\mu_p (2)^{\gamma}B^{\gamma+1}\frac{\Gamma(\gamma+n+2)}{\Gamma(n+1)}(\pi\alpha')^2
c_ic_X^*N^\mu R^{p-3}\nn\\
&&\times \bigg(\frac{\Lambda}{q}\bigg)^{\gamma+\frac{5}{2}}
x^{\frac{\gamma}{2}+n+\frac{9}{4}}(1-x)^{\frac{\gamma}{2}-\frac{1}{4}}\,.
\end{eqnarray}

\newpage

%-----------------------------------------------------------------------
\subsection{Results for the structure functions}
%-----------------------------------------------------------------------

In order to obtain $\textmd{Im}\:T^{\mu\nu}$, we multiply
Eq.(\ref{matrixelnonab}) by its complex conjugate and sum over the
radial excitations and over the polarizations of the final hadronic
states $\zeta_{X}^{\mu}$. The density of states can be estimated as
for the scalar and vector mesons. We then sum over polarizations and
neglect terms proportional to $q_{\mu}$ and $q_{\nu}$ as in the
previous section, obtaining
\be\label{kru47}
\textmd{Im} \:T^{\mu\nu} = C_f^2 \, \delta_{{\mathcal{AB_X}}} \,
\frac{\pi f f^*}{\Lambda s^{1/2}} \, \sum_{\lambda}N^{\mu}N^{*\nu}
= C_f^2 \, \delta_{{\mathcal{AB_X}}} \, \frac{\pi \, x^{\frac{1}{2}}
f f^*}{\Lambda
q(1-x)^{\frac{1}{2}}}(H_{\mu\nu}^{S}+H_{\mu\nu}^{A})\,,
\ee
where we have defined $\langle P+q, X|J^{\mu}(0)|P, {\cal
{Q}}\rangle = C_f \, \delta_{{\mathcal{AB}}_X} \,
f^{(\gamma)}_{\Lambda}(x,q) \, N^\mu$, while $H_{\mu\nu}^{S}$ and
$H_{\mu\nu}^{A}$ are exactly the same as Eqs. (\ref{kru50}) and
(\ref{kru51}).

By rewriting the hadronic tensor for spin-1 hadrons $W_{\mu\nu}$
from Eq.(\ref{DIS16}), we obtain the following structure functions
\begin{eqnarray}\label{structurefs}
F_{1}&=& A^{MF}(x)\frac{1}{12x^{3}}(1-x-2xt-4x^{2}t+4x^{3}t+8x^{3}t^{2})\,,\nn\\
F_{2}&=& A^{MF}(x)\frac{1}{6x^{3}}(1-x+12xt-14x^{2}t-12x^{2}t^{2})\,,\nn\\
b_{1}&=& A^{MF}(x)\frac{1}{4x^{3}}(1-x-xt)\,,\nn\\
b_{2}&=& A^{MF}(x)\frac{1}{2x^{3}}(1-x-x^{2}t)\,,\nn\\
b_{3}&=& A^{MF}(x)\frac{1}{24x^{3}}(1-4x+8x^{2}t)\,,\\
b_{4}&=& A^{MF}(x)\frac{1}{12x^{3}}(-1+4x-2x^{2}t)\,,\nn\\
g_{1}&=& A^{MF}(x)\frac{t}{8x^2}(-7+6x+8xt)\,,\nn\\
g_{2}&=& A^{MF}(x)\frac{1}{16x^{4}}(3-3x-4xt+2x^{2}t)\,,\nn
\end{eqnarray}
where
\begin{equation}
A^{MF}(x) = A^{MF}_0 \mu_p^2{\mathcal{Q}}^2 \alpha'^4 R^{2p-6}
\bigg(\frac{\Lambda^2}{q^2}\bigg)^{\gamma} x^{\gamma+2n+5}
(1-x)^{\gamma-1} \, ,
\end{equation}
and $A^{MF}_0 = 2 C_f^2 \, \delta_{{\mathcal{AB}}_X} \,
(2B)^{2\gamma+2} \, \pi^5 \,
\frac{\Gamma(\gamma+n+2)^2}{\Gamma(n+1)^2}
 \, |c_i^{({\mathcal{A}})}|^2| \, c_X^{({\mathcal{B}}_X)}|^2 \,$
is a dimensionless normalization constant. Subindex ${\mathcal{A}}$
labels the flavor of the incoming meson state, and ${\mathcal{B}}_X$
that of the intermediate state. These equations have been obtained
in the limit $|t|<<1$.

Notice that if we take the Abelian (single-flavored) limit we obtain
the same full set of structure functions calculated in section
\ref{ScalarVector}. Some particular cases have been calculated in
\cite{Koile:2011aa} and in appendix B\footnote{We can redefine the
normalization constants as $c_i^{{\mathcal(A)}}=c_i/\sqrt{C_f}$ and
$c_X^{({\mathcal{B}}_X)}=c_X/\sqrt{C_f}$ for the Abelian case.}.
Thus, we can summarize our results in a compact form
\be
F^{(a)MF}_i(x, t) = C_f^2 \, \delta_{{\mathcal{AB}}_X} \, F^{(a)SF}_i(x, t) \, ,
\ee
for each holographic dual model $(a)$, where $i$ indicates the each
particular structure function, $i=1, \cdot \cdot \cdot, 8$.

On the other hand, for each pair of holographic dual models $(a)$
and $(b)$ we find the relation $F^{(a)}_i(x, t) = A_{(a, b)}(x) \,
F^{(b)}_i(x, t)$, which leads to the following relation for the
hadronic tensor
\be
W^{\mu\nu}_{(a)}=A_{(a, b)}(x) \, W^{\mu\nu}_{(b)} \, .
\ee

Very interestingly, the set of Eqs.(\ref{structurefs}) leads to the
following inequality \footnote{These comments also hold for $N_f=1$,
see Eqs.(\ref{strfunctabel}).}
\begin{equation}\label{moements1}
 F_1 \geq |g_1| \, ,
\end{equation}
which holds for $|t| << 1$ for each Dp-brane model. This relation
implies the following inequality among moments of the
structure functions
\begin{equation}\label{moments3}
 M_n(F_1) \geq |M_n(g_1)| \quad n=1, 2, \ldots
\end{equation}
which must be satisfied from unitarity \cite{Manohar:1992tz}.
On the other hand, since $F_1\geq 0$ and $0\leq x\leq 1$, the chain of inequalities
\begin{equation}\label{moments2}
 M_n(F_1) \geq M_{n+1}(F_1) \quad n=1, 2, \ldots
\end{equation}
is satisfied. The moments of the structure functions $F_1$ and $g_1$ are defined as
follows
\ba
&& M_n(F_1) = \int_0^1 dx \, x^{n-1} \, F_1(x, q^2) \, ,
\\
&& M_n(g_1) = \int_0^1 dx \, x^{n-1} \, g_1(x, q^2) \, .
\ea

In addition, we have found relations between different structure
functions that we shall discuss in the conclusions.

%---------------------------------------------------------------------------
\section{Sub-leading contributions to the $1/N$ expansion}\label{LargeN}
%---------------------------------------------------------------------------

In this section we investigate the sub-leading contributions to the
$1/N$ and $N_f/N$ expansions of the two-point correlation functions
of global symmetry currents. We explain why in the large $N$ limit
we only have to consider the tree-level Witten's diagram displayed
in figure 1, which is the holographic dual version of the Feynman's
diagram of the forward Compton scattering of a charged lepton by a
hadron. We consider the full relevant Lagrangians of the three cases
studied in sections 3 and 4, corresponding to scalar mesons, $N_f=1$
vector mesons, and $N_f>1$ vector mesons, respectively. We study the
sub-leading contributions given by one-loop diagrams.

%--------------------------------------------------------------------------------
\subsection{Five-dimensional reduction of type IIB supergravity}
%--------------------------------------------------------------------------------

We very briefly review the five-dimensional reduction of type IIB
supergravity on $S^5$ as presented in \cite{Liu:1999kg} (other
relevant references for this section are
\cite{Freedman:1998tz,D'Hoker:1999jp,D'Hoker:1999pj}), in order to
give an example of the $N$-power counting in supergravity Feynman's
diagrams.

Let us begin with the ten-dimensional type IIB supergravity action
written in the Einstein frame, which contains the graviton, dilaton
$\phi$, the Ramond-Ramond axion field ${\cal {C}}$ and the five-form
field strength $F_5$
\be
S_{IIB}^{SUGRA} = - \frac{1}{2 \kappa_{10}^2} \, \int d^{10}x \,
\sqrt{|\det g|} \, \bigg[ {\cal {R}}_{10} - \frac{1}{2} \,
(\partial\phi)^2 -  \frac{1}{2} \, e^{2 \phi} \, (\partial {\cal
{C}})^2 - \frac{1}{4 \cdot 5!} \, (F_5)^2 \bigg] \, .
\ee
We consider the AdS$_5 \times S^5$ metric with the radius $R^4=4 \pi
g_s N \alpha'^2$.

Now, the five-dimensionally reduced action, in terms of the
five-dimensional dilaton $\phi_5(x)$ takes the form
\be
S_{5d}^{SUGRA} = - \frac{1}{2 \kappa_{5}^2} \, \int d^{5}x \,
\sqrt{|\det g_5|} \, \bigg[ {\cal {R}}_{5} - \frac{1}{2} \,
(\partial\phi_5)^2 + \cdot \cdot \cdot \bigg] \, ,
\ee
where dots indicate other terms which are not relevant for the
present discussion, since we are only interested in the $N$-power
counting. We consider the constant $\kappa_5$, which is defined as
\be
\frac{1}{2 \kappa_5^2} = \frac{N^2}{8 \pi^2} \, .
\ee
Hence, we can see how the factor $N^2$ appears in the
five-dimensional action. When we consider the D3D7-brane system, for
instance, the power-counting structure for the pure five-dimensional
supergravity action plus the DBI-action of the $N_f$ probe D7-branes
schematically reads
\be
S = N^2 \bigg[\tilde{S}^{SUGRA}_{IIB}+\frac{N_f}{N} \tilde{S}_{DBI}\bigg]
\, ,
\ee
where $\tilde{S}$ indicates the corresponding actions with kinetic
terms which do not depend on $N$. Thus, in order to obtain
canonically normalized fields we redefine the five-dimensional
dilaton as $\tilde\phi_5 \equiv N \phi_5$, and similarly for the
graviton. By plugging the normalized fields into the action $S$ one
obtains the correct power of $N$ in each interaction vertex.
Therefore, one can construct the Witten's diagrams for holographic
dual processes, displaying the corresponding $N$-power counting in
each case.

%---------------------------
\subsection{Scalar mesons}\label{scalarmesons}
%---------------------------

The relevant part of the free Lagrangian for scalar mesons
Eq.(\ref{scafree}) can be rewritten
as\footnote{We exclude the first term in Eq.(\ref{scafree}) since it does not contribute to the EOM.}
\begin{equation}\label{scafree2}
{\mathcal{L}}_{0}^{scalar}= -\mu_{p}(2\pi\alpha')^2 \, \sqrt{-\det
g} \, \bigg(\frac{\rho}{R}\bigg)^\beta \frac{1}{2}g^{ab}
\partial_a\Phi\partial_b\Phi^* \, .
\end{equation}
By factorizing the scalar field as
\be
\Phi(\rho, y^\mu, \Omega)=\sum_\ell \varphi^\ell(\rho, y^\mu) \, Y^\ell(\Omega) \,
,
\ee
and by defining the squared root of the determinant of the
five-dimensional piece of the metric as
\be
\sqrt{-\textmd{det}g^5} =
\bigg(\frac{\rho}{R}\bigg)^{\alpha+\frac{\beta}{2}} \, ,
\ee
we can write down the reduced five-dimensional free action for each $\varphi^\ell$ as
follows
\begin{equation}\label{scafree5d}
S_{0}^{scalar} = - \mu_{p} \, (\pi\alpha')^2 \, R^{p-4} \,
\int d^5x \, \sqrt{-{\det}g^5} \,
\bigg(\frac{\rho}{R}\bigg)^{(\alpha+\beta)+(p-4)
(1+\frac{\beta}{2})} \bigg[\partial^m\varphi^\ell \, \partial_m\varphi^{\ell*}
+ M_\ell^2 \, \varphi^\ell \, \varphi^{\ell*}\bigg] \, ,
\end{equation}
where we have used
\be\label{harmsph1}
\int d^{p-4}\Omega \, \sqrt{\textmd{det}\widetilde{g}} \,
Y^\ell \, Y^{\ell'*}=\delta_{\ell\ell'}\, ,
\ee
and we have defined
\be\label{harmsph2}
\int d^{p-4}\Omega \, \sqrt{\textmd{det}\widetilde{g}} \,
\eta^{ij} \, \partial_iY^\ell \, \partial_jY^{\ell'*}\equiv \delta_{\ell\ell'}M^2_\ell
\, .
\ee
Notice that Eq.(\ref{scafree5d}) is the action for a scalar complex
field.\footnote{Recall that we are using the signature (-,+,+,+,+).
The full action is $S_{0}^{scalar} = \sum_\ell S_{0}^{scalar \,
\ell}$, and we are writing only one $S_{0}^{scalar \, \ell}$ in
Eq.(\ref{scafree5d}).} Recall the interaction Lagrangian given in
Eq.(\ref{kru15})
\begin{eqnarray}
{\mathcal{L}}_{interaction}^{scalar}= i \, {\cal{Q}}\, \mu_{p} \,
(\pi\alpha')^{2} \, \sqrt{-\textmd{det}g} \,
\bigg(\frac{\rho}{R}\bigg)^{\beta} \, A^{m} \, (\Phi\partial_{m} \,
\Phi^{*}-\Phi^{*} \, \partial_{m}\Phi ) \, ,\nonumber
\end{eqnarray}
which under the same five-dimensional reduction becomes\footnote{
Notice that on the last equation we have dropped the subindex $X$
from the interaction Lagrangians. We keep this convention in the
rest of this section. Besides, we shall not write the superscript
$\ell$.}
\begin{equation}
S_{interaction}^{scalar}= i \, {\mathcal{Q}} \, \mu_{p} \, (\pi\alpha')^2 \, R^{p-4} \,
\int \, d^5x \, \sqrt{-{\det}g^5} \,
\bigg(\frac{\rho}{R}\bigg)^{(\alpha+\beta)+(p-4)(1+\frac{\beta}{2})}
\, A^m \, \bigg[\varphi \,
\partial_m\varphi^* - \varphi^* \, \partial_m\varphi\bigg] \, .
\end{equation}
At the order at which we are interested in, graviton-like
perturbations are relevant. By considering the $h_{mn}$ fluctuation
on the metric
\be\label{graviton}
g_{mn}\rightarrow g_{mn}+h_{mn} \, ,
\ee
this induces the interaction terms\footnote{The kinetic term for the
graviton as well as that for the gauge field $A^m$ come from the
ten-dimensional supergravity action discussed in the last
subsection. Here we only consider the $S_{DBI}$ discussed in
sections \ref{ScalarVector} and \ref{Non-Abelian}, defined in the
probe-brane worldvolume.} \label{kineticterms}
\be
h^{mn} \, \partial_m\Phi \, \partial_n\Phi^* \quad ; \quad h^{mn} \,
A_m \, \Phi \, \partial_n\Phi^*,
\ee
which, upon five-dimensional reduction of the action given in
Eq.(\ref{scafree5d}), become
\be
h^{mn} \, \partial_m\varphi \, \partial_n\varphi^* \quad , \quad
h^{mn} \, A_m \, \varphi \, \partial_n\varphi^*.
\ee
By assembling all factors, we obtain the full five-dimensional
action
\ba
S_{total}^{scalar}&=&S_{0}^{scalar}+S_{interaction}^{scalar}\nn\\
&=& -\mu_{p}(\pi\alpha')^2 R^{p-4} \, \int d^5x
\sqrt{-{\det}g^5}\bigg(\frac{\rho}{R}\bigg)^{(\alpha+\beta)+(p-4)(1+\frac{\beta}{2})}
\bigg[\partial^m\varphi\partial_m\varphi^* + M_\ell^2\varphi\varphi^*
+ \nn \\
& & h^{mn}\partial_m\varphi\partial_n\varphi^* - i{\mathcal{Q}} \,
A^m (\varphi\partial_m\varphi^*-\varphi^*\partial_m\varphi) -
i{\mathcal{Q}}h^{mn}A_m(\varphi\partial_n\varphi^*-\varphi^*\partial_n\varphi)\bigg]
, \nn \\
& &
\ea
which includes the kinetic term for the scalar mesons, as well as
the interaction terms with the graviton and the gauge field $A_m$.
The last two terms can be seen as part of a covariant derivative in
the kinetic term for the scalar meson. We can redefine the fields in
order to be canonically normalized
\ba
\widetilde{\varphi}\equiv\sqrt{N} \, \varphi \, ,\nn\\
\widetilde{A}^m\equiv N\, A^m \, ,\nn\\
\widetilde{h}^{mn}\equiv N \, h^{mn} \, .
\ea
By considering that
\be\label{globalfactor}
\mu_p(\pi\alpha')^2 =
\frac{(\pi\alpha')^2}{(2\pi)^p g_s \alpha'^{\frac{p+1}{2}}} =
\frac{1}{2^p\pi^{p-2}}\frac{N}{\lambda\alpha'^{\frac{p-3}{2}}},
\ee
we can write $S_{total}^{scalar}$ explicitly in terms having
different powers of $N$ as
\ba
S_{total}^{scalar}&=&\frac{R^{p-4}}{2^p\pi^{p-2}\lambda\alpha'^{\frac{p-3}{2}}}
\int d^5x\sqrt{-{\det}g^5}
\bigg(\frac{\rho}{R}\bigg)^{(\alpha+\beta)+(p-4)(1+\frac{\beta}{2})} \nn \\
& &
\bigg[\partial^m\widetilde{\varphi}\partial_m\widetilde{\varphi}^* +
M_\ell^2\widetilde{\varphi}\widetilde{\varphi}^* +
N^{-1}\widetilde{h}^{mn}\partial_m \widetilde{\varphi}\partial_n\widetilde{\varphi}^* - \nn\\
& & N^{-1} i {\mathcal{Q}} \widetilde{A}^m
(\widetilde{\varphi}\partial_m\widetilde{\varphi}^* -
\widetilde{\varphi}^*\partial_m\widetilde{\varphi}) - N^{-2} i
{\mathcal{Q}} \widetilde{h}^{mn} \widetilde{A}_m
(\widetilde{\varphi}\partial_n \widetilde{\varphi}^* -
\widetilde{\varphi}^*\partial_n\widetilde{\varphi}) \bigg].
\ea

Now, we can construct the relevant diagrams to our process as shown
in figures 1 and 2. We can see that only the tree-level diagram of
figure 1 contributes to leading order in $N$, namely, $N^{-2}$,
while diagrams of figure 2 are sub-leading, {\it i.e.} of order
$N^{-4}$.
\begin{figure}
\begin{center}
\epsfig{file=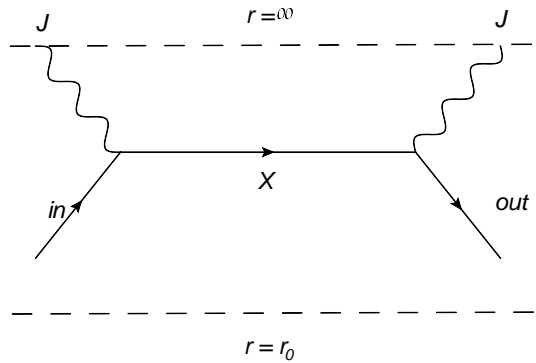, width=7cm} {\caption{\small
Five-dimensional tree-level Witten's diagram which is the
holographic dual version of the forward Compton scattering of a
charged lepton by a hadron in four-dimensions. This diagram is made
of five-dimensional fields obtained from dimensional reduction of
ten-dimensional supergravity on a compact Einstein manifold. The
dual field of the hadron is indicated with a solid line. The hadron
can be a scalar, a $N_f=1$ vector meson, or a $N_f>1$ vector meson.
Wavy lines indicate the dual field corresponding to a virtual photon
exchanged from the lepton (not drawn in this diagram) and the
hadron. This field (wavy line) is a fluctuation induced by the
insertion of the global symmetry current operator at the boundary
theory. This Witten's diagram gives the leading contribution to the
$1/N$ expansion. Notice that $r\rightarrow\infty$ corresponds to the
boundary where the four-dimensional gauge theory is defined and
where current operators $J$ are inserted.}} \label{Orden1}
\end{center}
\end{figure}
\begin{figure}
\begin{center}
\epsfig{file=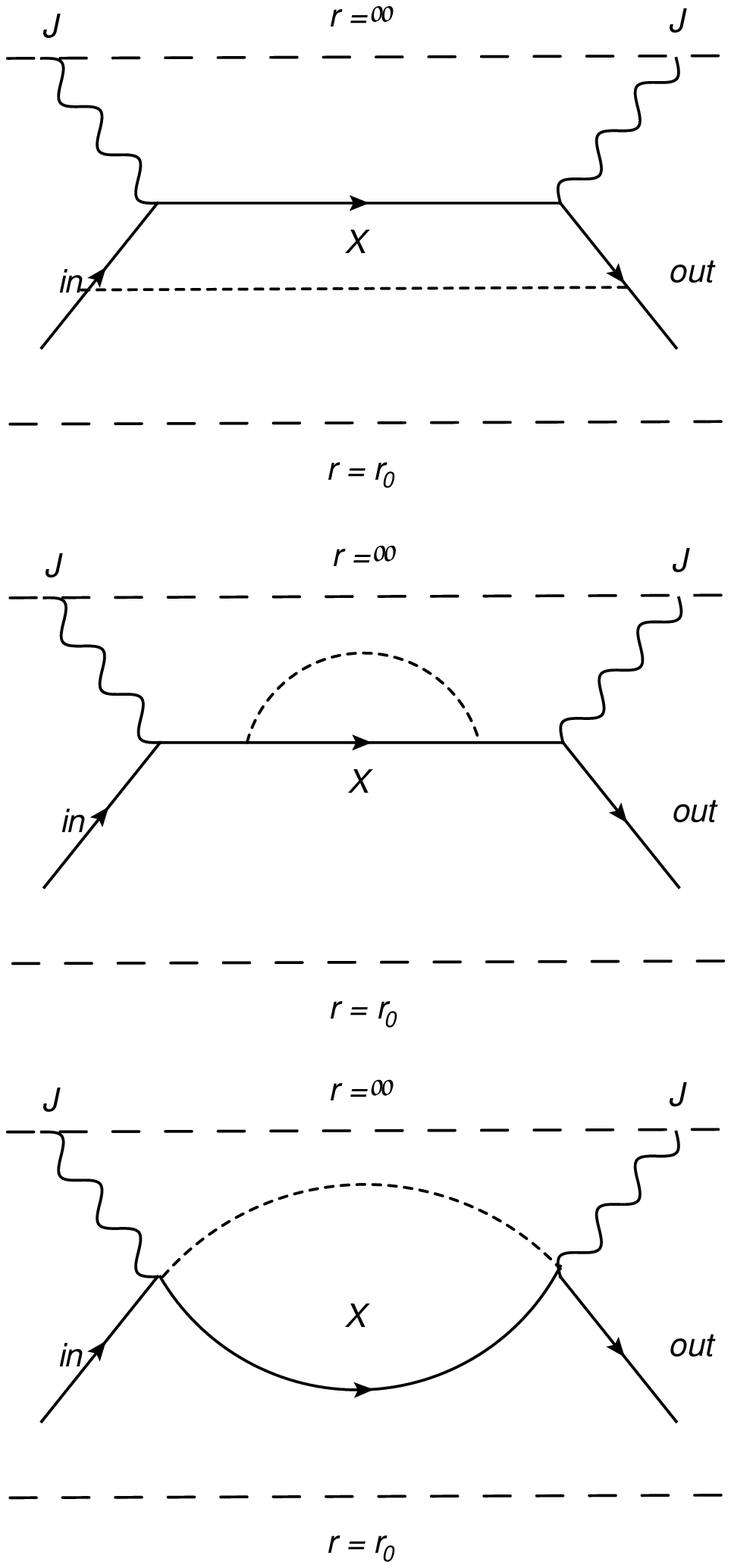, width=7cm} {\caption{\small
Illustration of some of one-loop Witten's diagrams. Five-dimensional
one-loop ladder graviton (top), rainbow graviton (middle), fish
graviton (bottom) Witten's diagrams corresponding to sub-leading
corrections to the forward Compton scattering Feynman's diagrams
contributing to DIS in four-dimensions. These diagrams are made of
five-dimensional fields obtained from dimensional reduction on a
five-dimensional Einstein manifold.  The dashed line indicates a
graviton $h_{mn}$. These Witten's diagrams contribute to order
$N^{-2}$ relative to the leading-order contribution in the $1/N$
expansion displayed in figure 1.}} \label{Orden2a}
\end{center}
\end{figure}

\newpage

%-------------------------------------------
\subsection{Vector mesons with $N_f=1$}\label{singleflavormesons}
%-------------------------------------------

Let us consider the relevant part of the free Lagrangian for $N_f=1$
vector mesons given in Eq.(\ref{kru41})
\begin{eqnarray}
{\mathcal{L}}_{0}^{SF} = - \mu_{p} \, (\pi \alpha')^{2} \,
\sqrt{-\det g} \,
 F^{ab} F_{ab}^{*} \, ,\nn
\end{eqnarray}
and define
\ba
B^a(\rho, y^\mu, \Omega) &=& \sum_\ell b^{a \, \ell}(\rho, y^\mu) \, Y^\ell(\Omega) \, , \nn\\
f_{mn}^{\ell} &=& \partial_mb_n^\ell - \partial_nb_m^\ell \, .
\ea
Then, we can write the reduced five-dimensional free action for each
$b^{\ell}_n$ as\footnote{From now on we drop the superscript $\ell$
in the rest of this subsection. The following equation is the part
of the action corresponding to only one $b^{\ell}_n$.}
\begin{equation}
S_{0}^{SF} = - \mu_{p} \, (2\pi\alpha')^2 R^{p-4} \, \int
d^5x \, \sqrt{-{\det}g^5} \,
\bigg(\frac{\rho}{R}\bigg)^{(\alpha+\beta)+(p-4)(1+\frac{\beta}{2})}
\bigg[\frac{1}{4}f^{mn}f_{mn}^* + \frac{1}{2}M_\ell^2 \, b^m b_m^*\bigg] \, ,
\end{equation}
which is a Proca-like Lagrangian.

After five-dimensional reduction the interaction Lagrangian
(\ref{kru42})
\begin{eqnarray}
{\mathcal{L}}_{interaction}^{SF} =i \, {\cal{Q}} \, \mu_{p} \, (\pi
\alpha')^{2} \, \sqrt{-\det g} \, A_{m} \, [B_{n}^{*} F^{nm} - B_{n}
(F^{nm})^{*}] \, , \nonumber
\end{eqnarray}
can be written as
\begin{equation}
S_{interaction}^{SF} = i \, {\mathcal{Q}} \, \mu_{p} \,
(\pi\alpha')^2 \, R^{p-4} \, \int d^5x \, \sqrt{-{\det}g^5}
\,
\bigg(\frac{\rho}{R}\bigg)^{(\alpha+\beta)+(p-4)(1+\frac{\beta}{2})}
\,  A^m \, \bigg[b_{n}^* f^{mn} - b_n f^{mn*}\bigg] \, .
\end{equation}
By considering a metric fluctuation, it introduces the interaction
terms
\be
h^{mq} \, F_q^n \, F_{mn}^* \quad , \quad h^{mq} \, A_m \, (B_{n}^*
\, F^n_q-B_{n}^* \, F_{q}^{n*}) \, ,
\ee
which, after dimensional reduction, become
\be
h^{mq} \, f_q^n f_{mn}^* \quad ; \quad h^{mq} \, A_m \, (b_{n}^*
f^n_q - b_{n}^* \, f_{q}^{n*}) \, ,
\ee
If we gather all these terms we obtain the full action
\ba
S_{total}^{SF}&=& S_{0}^{SF} + S_{interaction}^{SF} \nn \\
&=& -4\mu_{p}(\pi\alpha')^2R^{p-4} \, \int d^5x \sqrt{-{\det}g^5}
\bigg(\frac{\rho}{R}\bigg)^{(\alpha+\beta)+(p-4)(1+\frac{\beta}{2})}
\bigg[\frac{1}{4}f^{mn}f_{mn}^*+\frac{1}{2}M_\ell^2b^mb_m^* + \nn\\
& & \frac{1}{4}h^{mq} f_q^n f_{mn}^* - \frac{i}{4} {\mathcal{Q}}A^m \big(b_{n}^* f^{mn} -
b_n f^{mn*} \big) - \frac{i}{4} {\mathcal{Q}} h^{mq} A_m \big(b_{n}^* f^n_q -
b_{n}^* f_{q}^{n*} \big)\bigg].
\ea
We can redefine the fields in order to make the kinetic terms
canonically normalized in terms of powers of $N$, thus
\ba
\widetilde{b}^m &\equiv& \sqrt{N} \, b^m \, , \quad (\widetilde{f}^{mn}
\equiv \sqrt{N} \, f^{mn} \, )\nn \\
\widetilde{A}^m &\equiv& N \, A^m \, , \nn \\
\widetilde{h}^{mn} &\equiv& N \, h^{mn} \, .
\ea
By using Eq.(\ref{globalfactor}) we can write $S_{total}^{SF}$ in
terms of the powers of $N$ as
\ba S_{total}^{SF} &=&
\frac{R^{p-4}}{2^{p-2}\pi^{p-2}\lambda\alpha'^{\frac{p-3}{2}}} \int d^5x \sqrt{-{\det}g^5}
\bigg(\frac{\rho}{R}\bigg)^{(\alpha+\beta)+(p-4)(1+\frac{\beta}{2})} \nn \\
&& \bigg[\frac{1}{4}\widetilde{f}^{mn} \widetilde{f}_{mn}^* + \frac{1}{2} M_\ell^2
\widetilde{b}^m \widetilde{b}_m^* + N^{-1}\frac{1}{4} \widetilde{h}^{mq} \widetilde{f}_q^n
\widetilde{f}_{mn}^* - \nn \\
&& N^{-1}\frac{i}{4} {\mathcal{Q}} \widetilde{A}^m \big(\widetilde{b}_{n}^*
\widetilde{f}^{mn} - \widetilde{b}_n f^{mn*}\big) - N^{-2} \frac{i}{4}
{\mathcal{Q}} \widetilde{h}^{mq} \widetilde{A}_m
\big(\widetilde{b}_{n}^* \widetilde{f}^n_q - \widetilde{b}_{n}^*
\widetilde{f}_{q}^{n*}\big) \bigg]. \ea

The relevant diagrams to our process are very similar to the ones in
the case of scalar mesons. They are those in figures 1 and 2, just
noting that the meson line now corresponds to the vector meson $b^m$
instead of the scalar meson $\varphi$ in last subsection. We can see
again that only the tree-level diagram in figure 1 contributes to
leading order in $N$, namely, $N^{-2}$, while diagrams in figure 2
are sub-leading, {\it i.e.} order $N^{-4}$.

%------------------------------------------
\subsection{Vector mesons with $N_f>1$}\label{multipleflavormesons}
%------------------------------------------

We begin with the relevant part of the Lagrangian for vector mesons
with $N_f>1$ from Eq.(\ref{freenonabelian})
\begin{eqnarray}
{\mathcal{L}}_{0}^{MF}=-\mu_p(\pi\alpha')^2 \, \sqrt{-\det g} \,
Tr\bigg\{ \widehat{F}_{ab}\widehat{F}^{ab \, *} +
\bigg(i\widehat{F}_{ab}^*[B^a,B^b]+c.c.\bigg)-[B_a^*,B_b^*][B^a,B^b]\bigg\},\nn
\end{eqnarray}
with $\widehat{F}_{ab}=\partial_a B_b-\partial_{b} B_a=F_{ab}-i[B_a,B_b]$, and we define
\ba
B^a(\rho,y^\mu, \Omega)=\sum_\ell b^{a \, \ell}(\rho,y^\mu)Y^\ell(\Omega),\nn\\
\widehat{f}_{mn}^\ell=\partial_mb_n^\ell-\partial_nb_m^\ell=f_{ab}^\ell-i[b_a^\ell,b_b^\ell].
\ea
Then, we can write the reduced 5-dimensional free action
as\footnote{We write the piece of the action corresponding to a
single $b^\ell$.}
\begin{eqnarray}
S_{0}^{MF} & = & -2\mu_{p}(\pi\alpha')^2 R^{p-4} \,
\int d^5x\sqrt{-{\det}g^5}\bigg(\frac{\rho}{R}\bigg)^{(\alpha+\beta)+(p-4)(1+\frac{\beta}{2})}
Tr\bigg\{\frac{1}{2}\widehat{f}^{mn \, \ell}\widehat{f}_{mn}^{\ell \, *}
+M_\ell^2b^{m \, \ell}b_m^{\ell \, *}+ \nn\\ & &
\sum_{\ell' \, \ell''}\bigg[\frac{i}{2}a_{\ell \, \ell' \, \ell''}[b^{\ell'}_m,b^{\ell''}_n]
\widehat{f}^{mn \, \ell *}+c.c.\bigg]-\sum_{\ell' \, \ell'' \, \ell'''}
\frac{c_{\ell \, \ell' \, \ell'' \, \ell'''}}{2}[b_m^\ell,b_n^{\ell'}]
[b^{m \, \ell'' *},b^{n \, \ell''' *}]\bigg\},
\end{eqnarray}
where we have used Eqs.(\ref{harmsph1}) and (\ref{harmsph2}), and
defined
\ba
a_{\ell \, \ell' \, \ell''} & \equiv & \int d^{p-4}\Omega \, \sqrt{\widetilde{g}}
\, Y^{\ell*} \, Y^{\ell'} \, Y^{\ell''}, \\
c_{\ell \, \ell' \, \ell'' \, \ell'''} & \equiv & \int d^{p-4}\Omega \,
\sqrt{\widetilde{g}} \, Y^\ell \, Y^{\ell'} \, Y^{\ell''*} \, Y^{\ell'''*} .
\ea
After five-dimensional reduction the interaction Lagrangian
(\ref{kru42bis})
\begin{eqnarray}
{\mathcal{L}}_{interaction}^{MF}&=& i{\cal{Q}} \, \mu_{p} \, (\pi
\alpha')^{2} \, \sqrt{-\textmd{det}g} \, \bigg\{A_{m} \, Tr\big(B_{n}^{*} \, \widehat{F}^{nm}-B_{n} \, (\widehat{F}^{nm}\big)^{*})+\nonumber\\
&&\, i \, A_{m} \, Tr\big(B_{n}^* \, [B^{m} \, , \, B^{n}]\big) + i \, A_{m} \, Tr\big(B_n \, [B^{m*} \, , \, B^{n*}]\big) \,\bigg\}\ ,\\
&\equiv &{\mathcal{L}}_{interation_1}^{MF}+{\mathcal{L}}_{interation_2}^{MF}+{\mathcal{L}}_{interation_3}^{MF}.\nonumber
\end{eqnarray}
can be written as
\begin{eqnarray}
S_{interaction}^{MF} & = & {\mathcal{Q}}\mu_{p}(\pi\alpha')^2\int d^5x\sqrt{-{\det}g^5}
\bigg(\frac{\rho}{R}\bigg)^{(\alpha+\beta)+(p-4)(1+\frac{\beta}{2})}\\
& & \bigg\{ iA^mTr\bigg(b_{n}^{\ell \, *}\widehat{f}^{mn \, \ell}-b_n^\ell\widehat{f}^{mn \, \ell *}\bigg)-
\sum_{\ell' \, \ell''}\bigg(a_{\ell \, \ell' \, \ell''}A_mTr\bigg(b_{n}^{\ell \, *}
\big[b^{m \ell'},b^{n \ell''}\big]\bigg)+c.c.\bigg) \bigg\}.\nn
\end{eqnarray}

Graviton-like perturbations are relevant, thus it introduces the
interaction terms
\be
h^{mq}Tr(\widehat{F}_{q}^n\widehat{F}_{mn}^*) \quad ; \quad h^{mq}
Tr[A_m(B_{n}^*\widehat{F}^n_q-B_{n}^*\widehat{F}_{q}^{n*})],
\ee
which, after dimensional reduction, become
\be
h^{mq}Tr(\widehat{f}_{q}^n\widehat{f}_{mn}^*) \quad ; \quad h^{mq}
Tr[A_m(b_{n}^*\widehat{f}^n_q-b_{n}^*\widehat{f}_{q}^{n*})].
\ee
By assembling all factors, we obtain the full action
\ba
S_{total}^{MF}&=& S_{0}^{MF}+S_{interaction}^{MF}\nn\\ & = &
-2\mu_{p}(\pi\alpha')^2 R^{p-4}\int d^5x\sqrt{-{\det}g^5}
\bigg(\frac{\rho}{R}\bigg)^{(\alpha+\beta)+(p-4)(1+\frac{\beta}{2})}\nn\\ & &
Tr\bigg\{\frac{1}{2}\widehat{f}^{mn \, \ell}\widehat{f}_{mn}^{\ell \, *}+
M_\ell^2b^{m \, \ell}b_m^{\ell \, *}+
\sum_{\ell' \, \ell''}\bigg(\frac{i}{2}a_{\ell \, \ell' \, \ell''}
[b^{\ell'}_m,b^{\ell''}_n]\widehat{f}^{mn \, \ell *}+c.c.\bigg)\nn\\ & &
-\sum_{\ell' \, \ell'' \, \ell'''}\frac{c_{\ell \, \ell' \, \ell'' \, \ell'''}}{2}
[b_m^\ell,b_n^{\ell'}][b^{m \, \ell'' *},b^{n \, \ell''' *}]\bigg\}+
\frac{1}{2}h^{mq}Tr(\widehat{f}_{q}^{n \, \ell}\widehat{f}_{mn}^{\ell \, *})\nn\\ & &
- \frac{i}{2}{\mathcal{Q}}A^mTr\bigg(b_{n \, \ell}^*\widehat{f}^{mn \, \ell}-
b_n^{\ell}\widehat{f}^{mn \, \ell \, *}\bigg)- \frac{i}{2}{\mathcal{Q}}h^{mq}A_m
Tr\bigg(b_{n}^{\ell \, *}\widehat{f}^{n \, \ell}_q-b_{n}^{\ell \, *}
\widehat{f}_{q}^{n \, \ell \, *}\bigg)+\nn\\ & &
\bigg[\frac{i}{2}\sum_{\ell' \, \ell''}a_{\ell \, \ell' \, \ell''}A_m
Tr\bigg(b_{n \, \ell}^*[b^{m \, \ell'},b^{n \, \ell''}]\bigg)+c.c.\bigg]\bigg\}.
\ea
As before, we redefine the fields in order to be canonically
normalized:
\ba
\widetilde{b}^m &\equiv& \sqrt{N} \, b^m \, ,
\quad (\widetilde{\widehat{f}}^{mn} \equiv \sqrt{N} \, \widehat{f}^{mn} \, )\nn \\
\widetilde{A}^m &\equiv& N A^m,\nn\\
\widetilde{h}^{mn} &\equiv& Nh^{mn} \, .
\ea
By using Eq.(\ref{globalfactor}), we can write $S_{total}^{MF}$ in
terms of the powers $1/N$ as
\ba
S_{total}^{MF}&=&\frac{R^{p-4}}{2^{p-1}\pi^{p-2}\lambda\alpha'^{\frac{p-3}{2}}}
\int d^5x\sqrt{-{\det}g^5}\bigg(\frac{\rho}{R}\bigg)^{(\alpha+\beta)+(p-4)(1+\frac{\beta}{2})}\nn\\ & &
Tr\bigg\{\frac{1}{2}\widetilde{\widehat{f}}^{mn \,
\ell}\widetilde{\widehat{f}}_{mn}^{\ell \, *}
+M_\ell^2\widetilde{b}^{m \, \ell}\widetilde{b}_m^{\ell \, *}+
N^{-\frac{1}{2}} \sum_{\ell' \, \ell''}\bigg(\frac{i}{2}a_{\ell \,
\ell' \,
\ell''}[\widetilde{b}^{\ell'}_m,\widetilde{b}^{\ell''}_n]\widetilde{\widehat{f}}^{mn
\, \ell *}+c.c.\bigg)\nn\\ & &
-N^{-1}\sum_{\ell' \, \ell'' \, \ell'''}\frac{c_{\ell \, \ell' \,
\ell'' \, \ell'''}}{2}
[\widetilde{b}_m^\ell,\widetilde{b}_n^{\ell'}][\widetilde{b}^{m \,
\ell'' *},\widetilde{b}^{n \, \ell''' *}]\bigg\}+
N^{-1}\frac{1}{2}\widetilde{h}^{mq}Tr(\widetilde{\widehat{f}}_{q}^{n
\, \ell} \widetilde{\widehat{f}}_{mn}^{\ell \, *})\nn\\ & &
-N^{-1}\frac{i}{2}{\mathcal{Q}}A^mTr\bigg(\widetilde{b}_{n \,
\ell}^*\widetilde{\widehat{f}}^{mn \,
\ell}-\widetilde{b}_n^{\ell}\widetilde{\widehat{f}}^{mn \, \ell \,
*}\bigg)-
N^{-2}\frac{i}{2}{\mathcal{Q}}\widetilde{h}^{mq}A_mTr\bigg(\widetilde{b}_{n}^{\ell
\, *}\widetilde{\widehat{f}}^{n \, \ell}_q-\widetilde{b}_{n}^{\ell
\, *}\widetilde{\widehat{f}}_{q}^{n \, \ell \, *}\bigg)+\nn\\ & &
N^{-\frac{3}{2}}\bigg[\frac{i}{2}\sum_{\ell' \, \ell''}a_{\ell \,
\ell' \, \ell''}A_m Tr\bigg(\widetilde{b}_{n \,
\ell}^*[\widetilde{b}^{m \, \ell'},\widetilde{b}^{n \,
\ell''}]\bigg)+c.c.\bigg]\bigg\}.
\ea

Since we have more vertices, we can construct more diagrams relevant
to our process. These are, in addition to the ones in figures 1 and
2, those in figure 3. These new diagrams are sub-leading in $N$ as
well but, since they have a meson loop, we have to sum over all the
different flavors, obtaining then a factor $N_f$. In addition,
notice that these three last diagrams, while sub-leading with
respect to the tree-level diagram of figure 1, are dominant with
respect to those of figures 2, 3 and 4. The suppression of the
diagrams of figure 2 with respect of that of figure 1 is of order
$1/N^2$.
\begin{figure}
\begin{center}
\epsfig{file=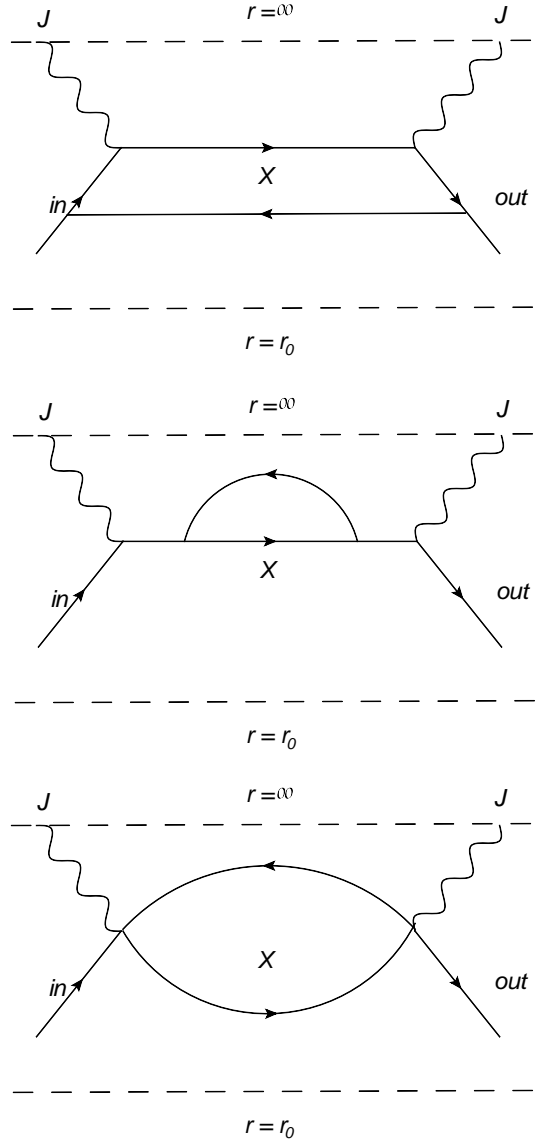, width=7cm} {\caption{\small
Illustration of some one-loop Witten's diagrams. Five-dimensional
one-loop ladder meson (top), rainbow meson (middle), fish-like meson
(bottom) Witten's diagrams corresponding to sub-leading corrections
to the forward Compton scattering Feynman's diagrams contributing to
DIS in four-dimensions. These diagrams are made of five-dimensional
fields obtained from dimensional reduction on a five-dimensional
Einstein manifold.  The solid line indicates a vector meson with
$N_f>1$. These Witten's diagrams contribute to order $N_f/N$
relative to the leading-order contribution in the $1/N$ expansion
displayed in figure 1.}} \label{Orden2a}
\end{center}
\end{figure}
%

%----------------------------------------------------------------------
\subsection{Higher-order contributions to the supergravity calculation}
%----------------------------------------------------------------------

In the previous subsections we have discussed the next-to-leading
order terms corresponding to the $1/N$ expansion from the scalar and
vector mesons. We have obtained those terms after re-scaling the
meson fields. The result is that one obtains more vertices in
comparison with the leading order Lagrangian. Therefore, we can
construct more diagrams which are relevant to the holographic dual
description of the forward Compton scattering process. The
additional diagrams are of the type presented in figures 2, 3 and 4.
In figure 2 we display three types of five-dimensional one-loop
Witten's diagrams: a ladder-graviton diagram (top), a
rainbow-graviton diagram (middle), and a fish-graviton diagram
(bottom). They correspond to sub-leading corrections to the forward
Compton scattering Feynman's diagrams contributing to DIS in
four-dimensions. Notice that these diagrams are made of
five-dimensional fields obtained from dimensional reduction of the
type IIA or IIB supergravity (depending on the model we consider) on
a five-dimensional Einstein manifold. The dashed line indicates a
graviton $h_{mn}$. These Witten's diagrams contribute to order
$N^{-2}$ relative to the leading-order contribution in the $1/N$
expansion displayed in figure 1. Notice that it can be additional
one-loop and multi-loop diagrams to the ones indicated in the
figures, however the present analysis of their contributions to the
$1/N$ and $N_f/N$ expansions will be valid.

In figure 3 we display five-dimensional one-loop ladder meson (top),
rainbow meson (middle), fish-like meson (bottom) Witten's diagrams
corresponding to sub-leading corrections to the forward Compton
scattering Feynman's diagrams contributing to DIS in
four-dimensions. As in figure 2 these diagrams are made of
five-dimensional fields obtained from dimensional reduction of
ten-dimensional supergravity on a five-dimensional Einstein
manifold. The solid line in the top figure indicates a vector meson with $N_f>1$.
Thus, these diagrams are sub-leading in the $1/N$ expansion, but
since they have a meson loop, we have to sum over all different
flavors, obtaining a factor $N_f$.

Therefore, the suppression of the diagrams in figure 2 is of order
$1/N^2$ while the suppression of diagrams in figure 3 is of the
order $N_f/N$.

In addition, we can also consider multi-loop contributions as in
figure 4. Five-dimensional two-loop meson (top), $n$-loop ladder
graviton (middle), $n$-loop rainbow graviton (bottom)  Witten's
diagrams corresponding to sub-leading corrections to the forward
Compton scattering Feynman's diagrams contributing to DIS in
four-dimensions. These diagrams are made of five-dimensional fields
obtained from dimensional reduction on a five-dimensional Einstein
manifold.  The solid line in the top diagram indicates a vector
meson with $N_f>1$. These Witten's diagrams contribute to order
$(N_f/N)^2$ (top) and $(1/N^2)^n$ (middle and bottom) relative to
the leading-order contribution in the $1/N$ expansion displayed in
figure 1.
\begin{figure}
\begin{center}
\epsfig{file=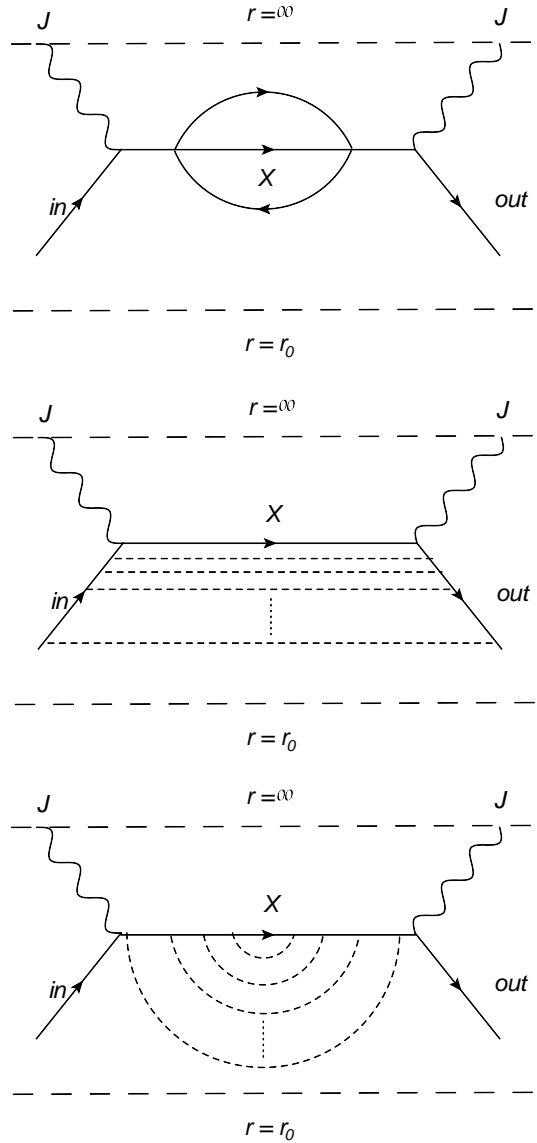, width=7cm} {\caption{\small
Illustration of some multi-loop Witten's diagrams. Multi-loop
contributions from Witten's diagrams corresponding to sub-leading
corrections to the forward Compton scattering Feynman's diagrams
contributing to DIS in four-dimensions.}} \label{Orden3a}
\end{center}
\end{figure}
Therefore, we can summarize the contributions as:
\begin{itemize}
\item Figure 1: Leading contribution.
\item Figure 2: Sub-leading contributions, of order $N^{-2}$ relative to the one in figure 1.
\item Figure 3: Sub-leading contributions, of order $N_f/N$ relative to the one in figure 1.
\item Figure 4: Sub-leading contributions, of order $(N_f/N)^2$ and $(1/N^2)^n$
relative to the one in figure 1.
\end{itemize}
We have not explicitly obtained these sub-leading contributions from
the diagrams illustrated in figures 2-4. Notice that the UV
completion of these diagrams should be done in terms of string
theory calculations.

%----------------------------------------------------------------------------
\subsection{Comments on the quantum field theory OPE}
%----------------------------------------------------------------------------

In this section we aim at relating the $1/N$ and $N_f/N$ expansions
discussed in the previous subsection from the supergravity point of
view with the corresponding expansions from the operator product
expansion of two-currents in the four-dimensional dual gauge
theories. First notice that the $T_{\mu\nu}$ tensor, whose
expectation value enters the definition of the hadronic tensor
$W_{\mu\nu}$, is given by the product of two currents
\be
\hat T_{\mu\nu} \equiv i \, \int d^4x \, e^{i q \cdot x} {\hat
{T}}(\hat J_\mu(x) \, \hat J_\nu(0)) \, . \nonumber
\ee
For deep inelastic scattering the leading operators in the OPE of
two currents are twist two when the gauge theory is weakly coupled.
So, to zeroth order in QCD one can write\footnote{This expression of
the OPE follows the notation and metric convention of
\cite{Hoodbhoy:1988am}, which has an overall minus sign in the
metric.}
\ba
\hat T_{\mu\nu} &=& \sum_{n=2,4,\cdot\cdot\cdot}^\infty C^{(1)}_n \,
\bigg( -g_{\mu\nu} + \frac{q_\mu q_\nu}{q^2} \bigg) \, \frac{2^n
q_{\mu_1} \cdot \cdot \cdot q_{\mu_n}}{(-q^2)^n} \, \hat O_V^{\mu_1
\cdot \cdot \cdot \mu_n} \nonumber \\
&+& \sum_{n=2,4,\cdot\cdot\cdot}^\infty C^{(2)}_n \, \bigg(
g_{\mu\mu_1} - \frac{q_\mu q_{\mu_1}}{q^2} \bigg) \bigg(
g_{\nu\mu_2} - \frac{q_\nu q_{\mu_2}}{q^2} \bigg) \, \frac{2^n
q_{\mu_3} \cdot \cdot \cdot q_{\mu_n}}{(-q^2)^{n-1}} \, \hat
O_V^{\mu_1 \cdot \cdot \cdot \mu_n} \nonumber \\
&+& \sum_{n=1,3,\cdot\cdot\cdot}^\infty C^{(3)}_n \, i \,
\epsilon_{\mu\nu\lambda\mu_1} \, q^\lambda \, \frac{2^n q_{\mu_2}
\cdot \cdot \cdot q_{\mu_n}}{(-q^2)^n} \, \hat O_A^{\mu_1 \cdot
\cdot \cdot \mu_n} \,  ,
\ea
where $C^{(1)}_n=C^{(2)}_n=C^{(3)}_n=1+{\cal {O}}(\alpha_s)$, where
$\alpha_s$ is the QCD coupling. The operators are defined as follows
\ba
&& \hat O_V^{\mu_1 \cdot \cdot \cdot \mu_n} = \frac{1}{2} \, \bigg(
\frac{i}{2} \bigg)^{n-1} \, \hat S\bigg({\bar\psi} \, \gamma^\mu \,
\hat D^{\mu_1} \cdot \cdot \cdot \hat D^{\mu_n} \, \hat {\cal
{Q}}_{qcm}^2
\, \psi \bigg) \, , \\
&& \hat O_A^{\mu_1 \cdot \cdot \cdot \mu_n}= \frac{1}{2} \, \bigg(
\frac{i}{2} \bigg)^{n-1} \, \hat S\bigg({\bar\psi} \, \gamma^\mu \,
\hat D^{\mu_1} \cdot \cdot \cdot \hat D^{\mu_n} \, \gamma_5 \, \hat
{\cal {Q}}_{qcm}^2 \, \psi \bigg) \, ,
\ea
where the derivative operator $\hat D^\mu$ acts on left and right,
while $\hat {\cal {Q}}_{qcm}$ is quark charge matrix, $\hat S$
indicates symmetrization and it removes all traces over $\mu_1 \cdot
\cdot \cdot \mu_n$.

Now, in order to calculate the structure functions in this regime
one has to calculate the matrix element of $T_{\mu\nu}$ between two
hadronic states, in the present case of spin-1, which leads to
\ba
&& <P, E|\hat O_V^{\mu_1 \cdot \cdot \cdot \mu_n}|P, E> = \hat S[a_n
\, P^{\mu_1} \cdot \cdot \cdot P^{\mu_n} + d_n \, \bigg(E^{*\mu_1}
E^{\mu_2}-\frac{1}{3} \, P^{\mu_1} \, P^{\mu_2}\bigg) \, P^{\mu_3}
\cdot \cdot \cdot P^{\mu_n}] \, , \nonumber \\
&& \\
&& <P, E|\hat O_A^{\mu_1 \cdot \cdot \cdot \mu_n}|P, E> = \hat S[r_n
\, \epsilon^{\lambda\sigma\tau\mu_1} \, E_\lambda^* \, E_\sigma \,
P_r \, P^{\mu_2} \cdot \cdot \cdot P^{\mu_n}]\, , \nonumber \\
&&
\ea
which define the coefficients $a_n$, $d_n$ and $r_n$.

The leading diagram in the parton model is displayed in figure 5
where the virtual photon strikes a parton. This is a tree-level
perturbative QFT calculation in the weakly coupled theory.
\begin{figure}
\begin{center}
\epsfig{file=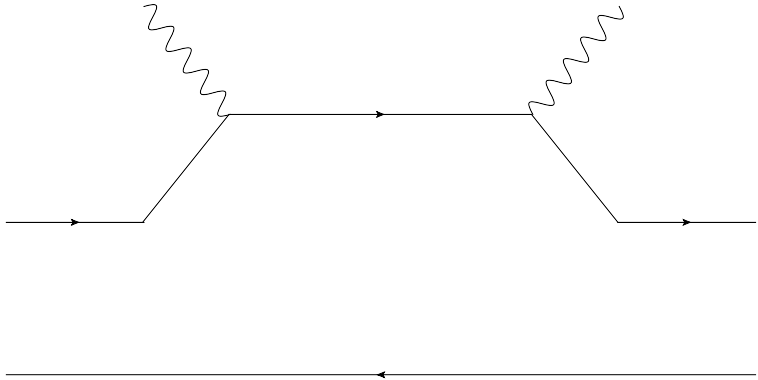, width=10cm} {\caption{\small Forward Compton
scattering from a meson in the parton model. A parton is struck by
the virtual photon indicated with a wavy line.}} \label{QCDparton}
\end{center}
\end{figure}
Thus, the operators which appear in the $JJ$ OPE at weak coupling
have twist $\tau =2, 4, \cdot \cdot \cdot$, even, and therefore
twist-two single-trace operators dominates the OPE. Notice that at
finite coupling these operators develop anomalous dimensions
$\gamma_n$, where the subindex stands for the quantum numbers of the
corresponding operator. In leading perturbation theory $\gamma_n
\sim \alpha_s(q^2) \, N$, and in this regime the parton model for
spin-$1/2$ partons leads to the Callan-Gross relation $F_2=2 x F_1$,
where the Bjorken variable $x$ is the fraction of the total momentum
($P^\mu$) of the hadron carried by the specified parton. The idea is
that a parton evolves, which means that it splits into more partons
which leads to reduce the momentum carried by each individual
parton.

On the other hand, at large coupling the situation changes
dramatically because the above operators have large anomalous
dimensions and then they no longer dominate the OPE. The point is
that on general grounds there are double-trace operators which do
not receive large anomalous dimensions for any value of the 't Hooft
coupling. It turns out that these operators dominate the OPE at
strong coupling. They are protected operators. Basically, the
discussion is similar to that presented for the case of a theory
with adjoint fields, where leptons are scattered by glueballs
\cite{Polchinski:2002jw}, but now there are contributions from
fields in the fundamental representation of the gauge group, which
leads us to replace the factor $N$ by $\sqrt{N}$ wherever it
corresponds when considering fundamental fields instead of adjoint
ones. Also, there will be a $N_f$ factor coming from summing over
flavor loops. The first difference with respect to the weak coupling
situation is that the lepton cannot strike individual partons any
more, and instead it strikes the hole hadron. This can be
represented by a quark-gluon diagram as in figure 6, which
represents a multi-gluon exchange in a planar diagram which can be
calculated in terms of its dual tree-level Witten's diagram of
figure 1, and they are the calculations that we have presented in
sections 2 and 3.
\begin{figure}
\begin{center}
\epsfig{file=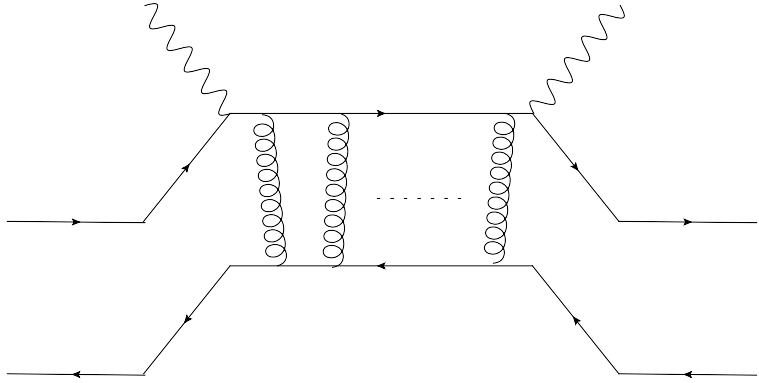, width=10cm} {\caption{\small Forward Compton
scattering at strong coupling. The meson is struck by the virtual
photon indicated with a wavy line. This is the quark-model diagram
which corresponds to the leading supergravity dual calculation given
by the tree-level Witten's diagram of figure 1. Multi-gluon exchange
between quark-anti quark pair is shown. This is a planar diagram.}}
\label{QCD-2}
\end{center}
\end{figure}
In principle, one can go beyond the planar limit and include
non-planar diagrams for gluon exchange as depicted in figure 7.
\begin{figure}
\begin{center}
\epsfig{file=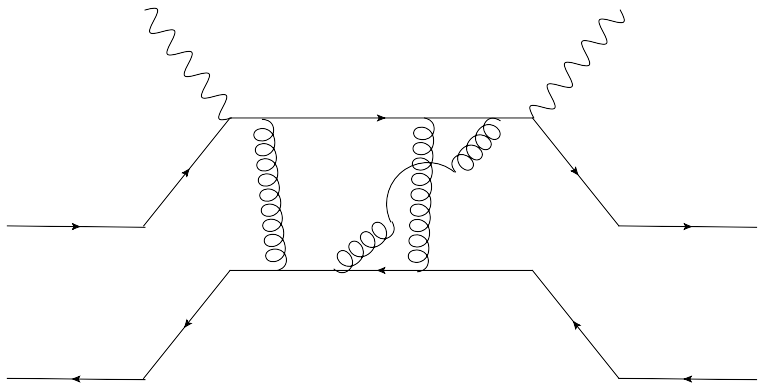, width=10cm} {\caption{\small Forward Compton
scattering at strong coupling. The meson is struck by the virtual
photon indicated with a wavy line. Non-planar multi-gluon exchange
between quark-anti quark pair is shown.}} \label{QCD-3}
\end{center}
\end{figure}
This is the quark-model diagram which corresponds to sub-leading
supergravity dual calculations of the type given by the one-loop
Witten's diagram of figure 2. Moreover, it is also possible to
consider multi-flavor loops as shown in figure 8.
\begin{figure}
\begin{center}
\epsfig{file=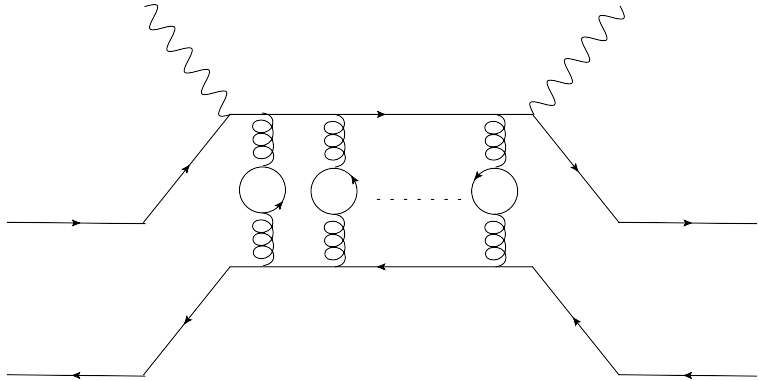, width=10cm} {\caption{\small Forward Compton
scattering at strong coupling. The meson is struck by the virtual
photon indicated with a wavy line. This is the quark-model diagram
which corresponds to sub-leading supergravity dual calculations of
the type given by the $n$-flavor loop Witten's diagram. These are
planar diagrams.}} \label{QCD-4}
\end{center}
\end{figure}
%

%----------------------------------------------------------------
\section{Discussion}\label{Discussion}
%----------------------------------------------------------------

As we mentioned in the introduction we have performed a detailed
analysis of the structure of the two-point correlation functions of
generic global symmetry currents at strong coupling, associated with
flavors in the fundamental representation of the gauge group, in the
quenched approximation, in terms of the corresponding holographic
string theory dual description. This includes the large $N$ limit of
supersymmetric and non-supersymmetric Yang-Mills theories in four
dimensions. In particular, we have explicitly investigated the cases
of the D3D7-brane, the D4D8$\mathrm{{\overline{D8}}}$-brane, and the
D4D6$\mathrm{{\overline{D6}}}$-brane systems.

In the large $N$ limit we have found a universal structure of the
two-point correlation functions of generic global symmetry currents
at strong coupling. For each holographic dual model we have found
that the two-point correlation functions of non-Abelian ($N_f > 1$)
global symmetry currents can generically be written as the product
of a constant, which depends on the particular Dp-brane model, times
flavor preserving Kronecker deltas multiplying the corresponding
Abelian ($N_f=1$) result for the same Dp-brane model. We have
obtained a universal factorization of the two-point correlation
functions for non-Abelian symmetry currents in a model-dependent
factor times a model-independent one. This has already been seen for
the two-point functions of Abelian symmetry currents in our previous
paper \cite{Koile:2011aa}. This factorization comes from the
structure of the flavored holographic dual model in the probe
approximation, where the probe Dp-brane action is taken to be the
non-Abelian version of the Dirac-Born-Infeld action
\cite{Tseytlin:1997csa}. Thus, in general we can write the hadronic
tensor $W^{\mu\nu}_{(a)}$ for a holographic dual model corresponding
to a certain gauge field theory in the large-$N$ limit as
\begin{equation}
W^{\mu\nu}_{(a)}=A_{(a, b)} \, W^{\mu\nu}_{(b)} \, ,
\end{equation}
for models $(a)$ and $(b)$, where $A_{(a, b)}(x)$ is a conversion
factor which depends on the pair of Dp-brane models considered. This
allows one to write the corresponding structure functions
$F_i^{(a)}(x, t)$, where subindex $i$ indicates the $i$-th structure
function for every meson in each particular model, as $F^{(a)}_i(x,
t) = A_{(a, b)} \, F^{(b)}_i(x, t)$ as we explained in the
Introduction. Besides, we have found that a modified version of the
Callan-Gross relation is satisfied for a large class of flavored
holographic dual models, $F_2 = 2 F_1$ without multiplying by the
Bjorken parameter, when the parameter $t \rightarrow 0$, which is an
indication that the coupling is strong and therefore there are no
partons. In fact, we have found a number of additional relations
among the structure functions which hold in every Dp-brane model
studied in the present context. They are
\ba
b_2 = 2 \,  b_1 \, , \\
b_1 = 3 \, F_1 \, , \\
g_2 = \frac{9}{4x} \,  F_1 \, \\
b_4 = -2 \, b_3 \, .
\ea
The last three relations are predictions as in our previous work
\cite{Koile:2011aa}. In addition, we have shown that all the moments
of the structure functions satisfy the corresponding inequalities
derived from unitarity, as expected \cite{Manohar:1992tz}.

These results, which hold for a number Dp-brane models, seem to
suggest that there is a universal structure of the two-point current
correlation functions, and therefore for the hadronic tensor.
Moreover, this might be an indication that the structure of actual
QCD polarized vector mesons at strong coupling should have the above
relations among their structure functions. QCD lattice calculations
could confirm these predictions, it would be very interesting to
know it. In the affirmative case, it would imply that any candidate
for a holographic QCD model in the large $N$ limit should lead to
two-point current correlation functions with the properties
indicated above.

On the other hand, it would be very interesting to know how the
above relations become modified at strong coupling for the
kinematical region where the Bjorken parameter is very small.
Additionally, it would also be extremely interesting to investigate
the fate of these new structure function relations at weak coupling.

A very interesting aspect of the present work is that we have
investigated the $1/N$ and $N_f/N$ contributions to the leading
order calculations of the hadronic tensor, from the supergravity
dual model point of view. Particularly, we have focused on the
structure of the relevant Lagrangians and Witten's diagrams. Indeed,
we have derived all relevant Lagrangians. On the other hand,
although we have not calculated these Witten's diagrams explicitly,
we have discussed how they arise from supergravity, how their $1/N$
and $N_f/N$ powers match those in the corresponding expansions in
quantum field theory, and how these Witten's diagrams are suppressed
by $1/N^2$ and $N_f/N$ powers, respectively, in the supergravity
dual models.

Other papers where holographic description of DIS has been
investigated include
\cite{Bayona:2009qe,BallonBayona:2010ae,Gao:2009ze,Gao:2010qk}.
However, we follow a different approach to construct the
interactions derived formally from the DBI action of the probe
branes in a way that explicitly manifests the global symmetry. Also,
other regimes of the Bjorken parameter have been considered through
their holographic dual description as for instance in references
\cite{Brower:2006ea,Cornalba:2008sp,Cornalba:2009ax,Cornalba:2010vk}.
In addition,  DIS and current correlators in SYM plasmas have also
been investigated \cite{Hatta:2007cs,CaronHuot:2006te}.
Particularly, $\alpha'^3$ type IIB string theory corrections to
current correlators in SYM plasmas have been investigated in
\cite{Hassanain:2009xw,
Hassanain:2011ce,Hassanain:2011fn,Hassanain:2010fv,Hassanain:2012uj}.

~

%%%%%%%%%%%%%%%%%%%%%%%%%%%%%%%%%%%%%%%%%%
\centerline{\large{\bf Acknowledgments}}
%%%%%%%%%%%%%%%%%%%%%%%%%%%%%%%%%%%%%%%%%%

~

We thank Horacio Falomir, Jos\'e Goity and Carlos N\'u\~nez for very
valuable comments and discussions. The work of E.K. and M.S. has
been partially supported by the ANPCyT-FONCyT Grant PICT-2007-00849,
and the CONICET-PIP-2010-0396 Grant. S.M. is supported by the
Department of Physics and Astronomy of Rutgers University.

\newpage

\appendix
%-----------------------------------------------------------------
\section{Appendix: Hadronic tensor of spin-one mesons}
%-----------------------------------------------------------------

In the definition of the hadronic tensor of spin-1 mesons in section
2 we have used the following functions
\begin{eqnarray}
\label{DIS17} && r_{\mu\nu}\equiv\frac{1}{(P \cdot q)^{2}}\bigg(q
\cdot \zeta^{*}\:q \cdot \zeta-\frac{1}{3}(P \cdot q)^{2}\kappa
\bigg)\eta_{\mu\nu} \, , \\
\label{DIS18} && s_{\mu\nu}\equiv\frac{2}{(P \cdot q)^{3}}\bigg(q
\cdot \zeta^{*}\:q \cdot \zeta-\frac{1}{3}(P \cdot q)^{2}\kappa
\bigg)P_{\mu}P_{\nu} \, , \\
\label{DIS19} && t_{\mu\nu}\equiv\frac{1}{2(P \cdot q)^{2}}\bigg(q
\cdot \zeta^{*}\:P_{\mu}\zeta_{\nu}+q \cdot
\zeta^{*}\:P_{\nu}\zeta_{\mu} +q \cdot
\zeta\:P_{\mu}\zeta^{*}_{\nu}+q \cdot \zeta
\:P_{\nu}\zeta^{*}_{\mu}-\frac{4}{3}(P \cdot q) P_{\mu}P_{\nu}\bigg) \, , \nonumber \\
&& \\
\label{DIS20} && u_{\mu\nu}\equiv\frac{1}{P \cdot
q}\bigg(\zeta^{*}_{\mu}\zeta_{\nu}+\zeta^{*}_{\nu}\zeta_{\mu}
+\frac{2}{3}M^{2}\eta_{\mu\nu}
-\frac{2}{3}P_{\mu}P_{\nu}\bigg) \, , \\
\label{DIS21} &&
s^{\sigma}\equiv\frac{-i}{M^{2}}\epsilon^{\sigma\alpha\beta\rho}
\zeta^{*}_{\alpha}\zeta_{\beta}P_{\rho} \, ,
\end{eqnarray}
being $\kappa=1 - 4 x^{2} t$ and $s^{\sigma}$ a four-vector
analogous to the spin four-vector in the case of spin-$\frac{1}{2}$
particles. Besides, $\zeta_\mu$ and $\zeta^*_\mu$ denote the initial
and final hadronic polarization vectors, respectively. The condition
$P \cdot \zeta=0$ is satisfied, and the normalization is given by
$\zeta^{2}=-M^{2}$.

~

%--------------------------------------------------------------------------------------------------------
\section{Appendix: Meson structure functions from the \\ D4D6$\mathrm{\mathbf{\overline{D6}}}$-brane model}
%--------------------------------------------------------------------------------------------------------

In this appendix we extend the calculations developed in
\cite{Koile:2011aa} to the model consisting of $N$ D4 and $N_f$ D6
branes described in \cite{Kruczenski:2003uq}. All the results
obtained in this section can be derived as particular cases of the
calculations in section \ref{ScalarVector}.\footnote{We have studied
the backgrounds D3D7-brane and D4D8$\mathrm{\overline{D8}}$-brane
systems in \cite{Koile:2011aa} the case $N_f=1$.} This model is
similar to that of \cite{Sakai:2004cn}, whose DIS calculations where
done in our previous paper \cite{Koile:2011aa}. Therefore, the
results will be similar.

The model in reference \cite{Kruczenski:2003uq} consists of branes
in the following configuration:
\ba\label{array}
N &   D4:   & 0  \  1  \  2  \  3  \  4 \   -  \  -    -    -    - \nonumber\\
N_f &   D6:   & 0  \  1  \  2  \  3    -  \  5  \ \ 6  \ \ 7  \  -    -.
\ea
Note that the D4 and the D6 branes may be separated from each other
along the directions $x_8$ and $x_9$. In the decoupling limit for
the D4-branes this system provides a non-conformal version of the
AdS/CFT correspondence. This means that on the gauge theory side
there is a supersymmetric five-dimensional $SU(N)$ gauge theory
coupled to a four-dimensional defect. The system is dual to
${\mathcal{N}}=2$ supersymmetric Yang-Mills theory in $d=4$. The
degrees of freedom localized on the defect are $N_f$ hypermultiplets
in the fundamental representation of $SU(N)$, which arise from the
open strings connecting the D4 and the D6-branes. Each
hypermultiplet consists on two Weyl fermions of opposite
chiralities, $\psi_L$ and $\psi_R$, and two complex scalars.

By identifying the direction  4 as $x_4\sim
x_4+\frac{2\pi}{M_{KK}}$, where $M_{KK}$ is the mass scale for the
Kaluza-Klein modes, and by imposing anti-periodic conditions for the
D4-brane fermions, all of the supersymmetries are broken and the
theory becomes a four-dimensional one for energies $E\ll M_{KK}$,
while the adjoint fermions and scalars become massive. Generation of
mass for the fundamental fermions is forbidden by a chiral $U(1)_A$
symmetry that rotates $\psi_L$ and $\psi_R$ with opposite phases.

In the  limit $N_f\ll N$, the back-reaction of the D6-branes on the
supergravity background is negligible, therefore, they can be
treated as probe branes. In the string description, the $U(1)_A$
symmetry corresponds to the rotation symmetry in the $89-$plane.

We adopt, as in \cite{Kruczenski:2003uq}, the solution in which
there are $N_f$ D6-branes and $N_f$ anti-D6-branes.

%------------------------------------------------------------------
\subsection*{Background of D4-branes}
%------------------------------------------------------------------

The background metric of $N$ D4-branes in this configuration is
\begin{equation}\label{embedd0}
ds^{2}=
\bigg(\frac{U}{R}\bigg)^\frac{3}{2}(\eta_{\mu\nu}dy^{\mu}dy^{\nu}+
f(U)d\tau^2)+\frac{(R^3U)^\frac{1}{2}}{\rho(U)^2}\overrightarrow{dz}\cdot
\overrightarrow{dz},
\end{equation}
with $U(\rho)=\Big(\rho^{3/2}+\frac{U_{KK}^3}{4\rho^{3/2}}\Big)$,
$f(U)=1-\frac{U^{3}_{KK}}{U^{3}}$,  and
$\overrightarrow{z}=(z^5,\ldots,z^9)$.

The dynamics of interest for DIS corresponds to the limit $q \gg
\Lambda$, where $\Lambda$ is the confinement energy scale of the
gauge theory. Thus, we shall consider the interaction in the UV
limit, being the interaction region given by $U_{int}\sim q^{2}
R^{3}\gg U_{0}= \Lambda^{2} R^{3} \equiv U_{KK}$. In this limit, the
induced metric on the D6-branes takes the form
\begin{equation}\label{embedd3}
ds^{2}=\bigg(\frac{U}{R}\bigg)^\frac{3}{2}\eta_{\mu\nu}dy^{\mu}dy^{\nu}+
\bigg(\frac{R}{U}\bigg)^\frac{3}{2}dU^{2}+R^\frac{3}{2}U^\frac{1}{2}
d\Omega_{2}^{2}\,,
\end{equation}
which is the same as Eq.(96) in \cite{Koile:2011aa}, coming from the
same limit ($U\gg U_{KK}$) taken in the context of the
$\textmd{D4D8}\mathrm{\overline{D8}}$-brane system model
\cite{Sakai:2004cn}. The difference is that the coordinates $z^8$
and $z^9$ do not belong to the probe brane in this case. We can see
that this metric is a particular case of (\ref{metgral}) with $p=6 \
; \ \alpha=-\beta=\frac{3}{2}$. Therefore, all the analysis done in
section \ref{ScalarVector} applies. We then write the main results,
avoiding further details.

%------------------------------------------------------------------
\subsection*{The gauge field}
%------------------------------------------------------------------

By proposing the {\it Ansatz} (\ref{gaugeansatz}) we obtain the
solution (\ref{gaugesol}) which reads ($\alpha=-\beta=\frac{3}{2} \
; \ p=6$)
\begin{eqnarray}\label{ss55}
A_{\mu}&=&\frac{2}{\Gamma(5/4)}n_{\mu}\:e^{iq \cdot
y}\bigg(\frac{q^2R^{3}}{U}\bigg)^{5/8}K_{5/4}\bigg(
\bigg[\frac{4q^2R^3}{U}\bigg]^\frac{1}{2}\bigg)\,,\\
A_{U}&=&-\frac{2i(q \cdot n)}{\Gamma(5/4)q}\frac{1}{(qR)^{3}}\:e^{iq
\cdot y}
\bigg(\frac{q^2R^{3}}{U}\bigg)^{17/8}K_{1/4}\bigg(
\bigg[\frac{4q^2R^3}{U}\bigg]^\frac{1}{2}\bigg)\,=-\frac{i}{q^{2}}\eta^{\mu\nu}q_{\mu}\partial_{U}A_{\nu}\,,
\end{eqnarray}
where $K_{5/4}$ and $K_{1/4}$ are modified Bessel functions, and $q
\equiv \sqrt{q^{2}}$.

%------------------------------------------------------------------
\subsection*{DIS from scalar mesons}
%------------------------------------------------------------------

The EOM for scalar mesons arises from the transversal fluctuation
\begin{equation}\label{scalarfluc}
z^{8}=0+2\pi \alpha'\chi\,,
\:\:\:\:\:\:\:\:\:\:\:\:\:\:\:z^{9}=0+2\pi\alpha'\varphi\,,
\end{equation}
where the coordinates $z^{8}$ and $z^{9}$ lie on the (8, 9) plane,
transversal to the D6-brane. Note that we are perturbing around the
solution $z^8=z^9=0$, \emph{i.e.} , $r=0$ which corresponds to the
$\textmd{D4D6}\overline{D6}$-brane system solution. The scalar
fluctuations around this background are $\chi$ and $\varphi$.

The Lagrangian for the scalar fluctuations at leading order is
\begin{equation}\label{leadingaction}
{\mathcal{L}}=-\mu_{6}\sqrt{|\textmd{det}{g}|}\bigg[1+\bigg(\frac{R}{U}\bigg)^{3/2}g^{ab}(\pi\alpha')^{2}
\partial_{a}\Phi \partial_{b} \Phi^*\bigg]\,,
\end{equation}
where $\Phi\equiv \chi+i\varphi$. The solutions are, after  imposing
an {\it Ansatz} like Eq.(\ref{kru12}),
\begin{equation}\label{ss71}
\Phi_{IN/OUT}=c_{i}(\Lambda U)^{-\frac{7}{8}-\frac{\gamma}{2}} e^{iP \cdot y}
Y(S^2)\,,
\end{equation}
\begin{equation}\label{ss72}
\Phi_{X}=c_{X} (s^{\frac{1}{4}}\Lambda^{-\frac{1}{2}})(\Lambda U)^{-\frac{7}{8}}
J_{\gamma}\bigg(\bigg[\frac{4sR^3}{U}\bigg]^\frac{1}{2}\bigg) e^{iP_{X} \cdot
y} Y(S^2)\,,
\end{equation}
with $\gamma=(1/4)\sqrt{49+64 \ell(\ell+1)}$. After perturbing the
metric as in (\ref{pol5}) we obtain
\begin{eqnarray}\label{ss68}
{\mathcal{L}}_{interaction}^{scalar \ D4D6}=i{\cal{Q}}\mu_{6}(\pi
\alpha')^{2}\sqrt{|\textmd{det} g|}\bigg(\frac{R}{U}\bigg)^\frac{3}{2}A_{m}(\Phi
\partial^{m}\Phi^{*}_{X}-\Phi^{*}_{X}\partial^{m}\Phi )\,={\cal{Q}}\sqrt{|\textmd{det}
g|}A_{m}j^{m},
\end{eqnarray}
with
\begin{equation}\label{ss69}
j^{m}=i\mu_{6}(\pi \alpha')^{2}\bigg(\frac{R}{U}\bigg)^\frac{3}{2}(\Phi
\partial^{m}\Phi^{*}_{X}-\Phi^{*}_{X}\partial^{m}\Phi )\,.
\end{equation}
By using the current conservation of Eq.(\ref{currentcons}) and the
{\it Ansatz} (\ref{kru21.1}), we obtain the structure functions
\begin{equation}\label{ss78}
F_{1}=0, \,\,\,\,\,\,\,\,\,\:\:\:\:F_{2}=A_{0\ D4D6}^{scalar}{\cal{Q}}^{2}
\bigg(\frac{\mu_{6}^{2}\alpha'^{4}}{\Lambda^{6}}\bigg)
\bigg(\frac{\Lambda^{2}}{q^{2}}\bigg)^{\gamma+1}x^{\gamma+7/2}(1-x)^{\gamma}\,,
\end{equation}
where $A_{0\
D4D6}^{scalar}=4\pi^{5}|c_{i}|^{2}|c_{X}|^{2}[\Gamma(9/4+\gamma)]^{2}[\Gamma(5/4)]^{-2}$
is a normalization dimensionless constant. We can see that this
solution is a particular case of Eq.(\ref{kru23}).

%----------------------------------------------------------------------
\subsection*{DIS from vector mesons}
%----------------------------------------------------------------------

From the DBI action we derive the EOM (\ref{vectexpand}), then we
propose a quadratic Lagrangian from which we obtain exactly the same
EOM. The {\it Ansatz} for the solution is (\ref{vectAns}) and the
solutions (\ref{solvectin}) and (\ref{solvectx}) become
\begin{eqnarray}\label{ss86}
B_{\mu \,IN/OUT}=\zeta_{\mu}c_{i}\Lambda^{-1}(\Lambda
U)^{-\gamma/2-7/8}e^{iP \cdot y}\:Y^{\ell}(S^{4})\,,
\end{eqnarray}
\begin{eqnarray}\label{ss87}
B_{X \mu}=\zeta_{X \mu}c_{X}\Lambda^{-1}(s^{-1/4}\Lambda^{-1/2})
\big(\frac{U}{\Lambda^2R^3}\big)^{-7/8}J_{\gamma}\bigg[\bigg(\frac{4sR^3}{U}\bigg)^\frac{1}{2}\bigg]e^{iP_{X}.y}\:Y^{\ell}(S^{2})\,,
\end{eqnarray}
with $\gamma=(1/4)\sqrt{49+64\ell(\ell+3)}$. The quadratic
Lagrangian is exactly that in Eq.(\ref{kru41}), the interaction
Lagrangian that in (\ref{kru42}), and by repeating the calculations
as in section \ref{ScalarVector}, we finally obtain the solutions
(\ref{strfunctabel}) with
\begin{eqnarray}
A_{D4D6}^{vect}(x) &=& A_{0\ D4D6}^{vect}{\cal{Q}}^{2}\bigg(\frac{\mu_{6}^{2}(\alpha')^{4}}{\Lambda^{6}}\bigg)
\bigg(\frac{\Lambda^{2}}{q^{2}}\bigg)^{\gamma}
x^{\gamma+11/2}(1-x)^{\gamma-1}\,,\\
A_{0\ D4D6}^{vect}&=&\pi^{5}|c_{i}|^{2}|c_{X}|^{2}[\Gamma(\gamma+9/4)]^{2}[\Gamma(5/4)]^{-2}.
\end{eqnarray}
We can observe that they have the same form as the ones obtained
with the D3D7 and D4D8$\mathrm{\overline{D8}}$-brane model studied
in our previous work \cite{Koile:2011aa}, and summarized in section
\ref{ScalarVector}.

\end{document}